\documentclass[]{emulateapj}
\usepackage{epsfig}
\usepackage{graphicx}
\usepackage{longtable}
\usepackage{rotating}

\begin{document}

\title{Discovery of kilohertz quasi-periodic oscillations and state
transitions in the low-mass X-ray binary 1E~1724--3045 (Terzan~2)}

\author{D. Altamirano\altaffilmark{1}, 
M. van der Klis\altaffilmark{1},
M.  M\'endez\altaffilmark{1,2},
R. Wijnands\altaffilmark{1}, \\
C. Markwardt \altaffilmark{3,4} \&
J. Swank \altaffilmark{3}\\
}

\email{d.altamirano@uva.nl}

\altaffiltext{1}{Astronomical Institute, ``Anton Pannekoek'',
  University of Amsterdam, and Center for High Energy Astrophysics,
  Kruislaan 403, 1098 SJ Amsterdam, The Netherlands.}
\altaffiltext{2}{Kapteyn Astronomical Institute, University of
  Groningen, P.O. Box 800, 9700 AV Groningen, The Netherlands.}
\altaffiltext{3}{Laboratory for High-Energy Astrophysics, NASA Goddard
  Space Flight Center, Greenbelt, MD 20771, U.S.A.}
\altaffiltext{4}{Department of Astronomy, University of Maryland,
  College Park, MD 20742., U.S.A.}

\begin{abstract}
We have studied the rapid X-ray time variability in 99 pointed
observations with the \textit{Rossi X-ray Timing Explorer} (RXTE)'s
Proportional Counter Array of the low-mass X-ray binary 1E~1724--3045
which includes, for the first time, observations of this source in its
island and banana states, confirming the atoll nature of this
source. We report the discovery of kilohertz quasi-periodic
oscillations (kHz QPOs). Although we have 5 detections of the lower kHz
QPO and one detection of the upper kHz QPO, in none of the
observations we detect both QPOs simultaneously.  By comparing the
dependence of the rms amplitude with energy of kHz QPOs in different
atoll sources, we conclude that this information cannot be use to
unambiguously identify the kilohertz QPOs as was previously
thought. We find that Terzan~2 in its different states shows 
timing behavior similar to that seen in other neutron-star low mass X-ray
binaries (LMXBs). We studied the flux transitions observed between
February 2004 and October 2005 and conclude that they are due to
changes in the accretion rate.

\end{abstract}
\date{}
\maketitle

\section{Introduction}
\label{sec:intro}

Low-mass X-ray binaries (LMXBs) can be divided into systems containing
a black hole candidate (BHC) and those containing a neutron star
(NS). The accretion process onto these compact objects can be studied
through the timing properties of the associated X-ray emission
\citep[see, e.g., ][for a review]{Vanderklis06}.  \citet{Hasinger89}
classified the NS LMXBs based on the correlated variations of the
X-ray spectral and rapid X-ray variability properties. They
distinguished two sub-types of NS LMXBs, the Z sources and the atoll
sources, whose names were inspired by the shapes of the tracks that
these sources trace out in an X-ray color-color diagram (CD) on time scales of
hours to days. The Z sources are the most luminous, but the atoll
sources are more numerous and cover a much wider range in luminosities
\citep[e.g. ][and references within]{Ford00}. For each type of source,
several spectral/timing states are identified which are thought to
arise from qualitatively different inner flow configurations
\citep{Vanderklis06}.  In the case of atoll sources, the three main
states are the extreme island state (EIS), the island state (IS) and
the banana branch, the latter subdivided into lower-left banana (LLB),
lower banana (LB) and upper banana (UB) states.
The EIS and the IS occupy the spectrally harder parts of the color
color diagram and correspond to lower levels of X-ray luminosity
($L_x$).  The associated patterns in the CD are traced out in hours to
weeks.  The hardest and lowest $L_x$ state is the EIS, which shows
strong \citep[up to 50\% rms amplitude, see][ and references
within]{Linares07} low-frequency flat-topped noise also known as
band-limited noise (BLN).  The IS is spectrally softer and has higher
X-ray luminosity than the EIS.  Its power spectra are characterized by
broad features and a dominant BLN component which becomes weaker and
generally higher in characteristic frequency as the flux increases and
the $>6$ keV spectrum gets softer.  In order of increasing $L_x$ we
then encounter the LLB, where twin kHz QPOs are generally first
observed, the LB, where 10-Hz BLN is still dominant and finally, the
UB, where the $<1$~Hz (power law) very low frequency noise (VLFN)
dominates.  In the banana states, some of the broad features observed
in the EIS and the IS become narrower (peaked) and occur at higher
frequency.
In particular, the twin kHz QPO can be found in the LLB at
frequencies higher than 700~Hz \citep[the lower of them with
frequencies generally lower than a 1000~Hz, while the upper one up to
1258~Hz, see e.g.,][]{Belloni05a,Jonker07}, only one kHz QPO can be
generally found in the LB, and neither of them is detected in the UB
\citep[see reviews by][]{Vanderklis00,Vanderklis04,Vanderklis06}.

A small number of weak NS LMXBs 
which do not get brighter than a few times $10^{36}$ ergs s$^{-1}$
\citep[usually burst sources and often referred to as 'weak' or 'faint
bursters', see e.g,][ and references within]{Muno03}
 resemble atoll sources in the EIS, but in the absence of state
 transitions this identification has been tentative \citep[see for
   example][]{Barret00a,Vanderklis06}.  An important clue is provided
 by the correlations between the component frequencies (and strengths
 - see e.g. \citealt{Straaten02,Straaten03,Altamirano08}) which helps
 to identify components across sources.
For example, \citet{Straaten02,Straaten03} compared the timing
properties of the atoll sources 4U~0614+09, 4U~1608--52 and
4U~1728--34 
(see also \citealt{Altamirano08}, for similar results when
the atoll source 4U~1636--53 was included in the sample)
and conclude that the frequencies of the variability
components in these sources follow the same pattern of correlations
when plotted versus the frequency of the upper kHz QPO ($\nu_u$).
\citet{Straaten03} also showed that low luminosity systems
extend the frequency correlations observed for the atoll sources. This
last result gave further clues in the link between the atoll and the
low luminosity sources.

\citet{Psaltis99} found an approximate frequency correlation involving a
 low-frequency QPO, 
 the lower kHz QPO frequency and two broad noise components
interpreted as low-frequency versions of these features.  This
correlation spans nearly three decades in frequency, where the Z and
bright atoll sources populate the $>100$\,Hz range and black holes and
weak NS systems the $<10$\,Hz range. As already noted by
\citet{Psaltis99}, because the correlation combines features from
different sources which show either peaked  or broad components with
relatively little overlap, the data are suggestive but not
conclusive with respect to the existence of a single correlation
covering this wide frequency range \citep{Vanderklis06}.

The low-luminosity neutron star systems can play a crucial role in
clearing up this issue. Observations of different source states in
such a system could connect the $<10$ and $>100$~Hz regions mentioned
above by direct observation of a transition in a single source.  In
the case of the pattern of correlations reported by
\citet{Straaten03}, low luminosity NS systems extend the frequency
correlations observed for ordinary atoll sources down to $\sim100$~Hz.
Unfortunately, the low luminosity NS systems are usually observed in
only one state (EIS), which makes it difficult to properly link these
sources to the atoll sources. However, some of these objects show rare
excursions to higher luminosity levels which might correspond to other
states. The occurrence of these excursions are usually
unpredictable. Therefore, in practice it was not possible until now to
check on the frequency behavior of the different variability
components as such a source enters higher luminosity states.

1E~1724--3045 is a classic low luminosity LMXB; a persistent Low-Mass
X-ray binary located in the globular cluster Terzan~2
\citep{Grindlay80} which is a metal-rich globular cluster of the
galactic bulge.  Its distance is estimated to be between 5.2 to 7.7
kpc \citep{Ortolani97}.  These values are consistent with that derived
from a type I X-ray burst that showed photospheric expansion
\citep[see][ but also see
\citealt{Kuulkers03,Galloway06}]{Grindlay80}.  The type I X-ray bursts
observed from this source also indicate that the compact object is a
weakly magnetized neutron star
\citep{Swank77,Grindlay80}. \citet{Emelyanov02} have shown, using
$\sim30$ years of data from several X-ray satellites, that the
luminosity of Terzan~2 increased until reaching a peak in 1997, after
which it started to decrease. They suggest that the evolution of the
donor star or the influence of a third star could be the cause of this
behavior. \citet{Olive98} and \citet{Barret00} have shown that during
earlier observations of Terzan~2 its X-ray variability at 
frequencies $\gtrsim0.1$~Hz resembled that of black hole
candidates. This state was tentatively identified as the extreme
island state for atoll sources. Until now, no kilohertz quasi-periodic
oscillations have been reported for this source, which was attributed
to the fact that the source was always observed in a single intensity
state \citep{Barret00a}.

Monitoring observations by the All Sky Monitor aboard the Rossi X-ray
Timing Explorer showed that the source was weakly variable in X-rays
(less than about a factor of 3 on a few day time scale for the first
8 years of the monitoring).  However, recently \citet{Markwardt04} 
reported (using PCA monitoring observations of the galactic bulge -
\citealt{Swank01}) that during February 2004, 1E~1724--3045 flared up
from its relatively steady $\sim20$ mCrab to $\sim66$~mCrab
(2--10~keV).
In this paper we report a complete study of the timing variability of
the source.
For simplicity, and since only one bright X-ray source is detected in the
globular cluster (See Section~\ref{sec:samesource}), in the rest of
this report we will refer to 1E~1724--3045 as Terzan~2.

\section{OBSERVATIONS AND DATA ANALYSIS}
\label{sec:dataanalysis}

\subsection{Light curves and color diagrams}\label{sec:lcandccd}
We use data from the Rossi X-ray Timing Explorer (RXTE) Proportional
Counter Array \citep[PCA; for instrument information
see][]{Zhang93,Jahoda06}. There were 534 slew observations until
October $30^{th}$ 2006, which are part of the PCA monitoring
observations of the galactic bulge \citep{Swank01} and which were
performed typically every 3 days. These observations were only used to
study the long-term $L_x$ behavior of the source.

There were also 99 pointed observations in the nine data sets we used
(10090-01, 20170-05, 30057-03, 50060-05, 60034-02, 80105-10, 80138-06,
90058-06 \& 91050-07), containing $\sim0.8$ to $\sim26$ ksec of useful
data per observation.  We use the 16-s time-resolution Standard 2 mode
data to calculate X-ray colors.  Hard and soft color are defined as
the 9.7--16.0 keV / 6.0--9.7 keV and 3.5--6.0 keV / 2.0--3.5 keV count
rate ratio, respectively, and intensity as the 2.0--16.0 keV count
rate. The energy-channel conversion was done using the pca\_e2c\_e05v02
table provided by the RXTE Team\footnote{see
http://heasarc.gsfc.nasa.gov/docs/xte/pca\_news.html}. Channels were
linearly interpolated to approximate these precise energy limits.
X-ray type I bursts were removed, background was subtracted and
deadtime corrections were made.  In order to correct for the gain
changes as well as the differences in effective area between the PCUs
themselves, we normalized our colors by the corresponding Crab Nebula
color values; \cite[see][ see table 2 in \citealt{Altamirano08} for
average colors of the Crab Nebula per
PCU]{Kuulkers94,Straaten03} that are closest in time but in the same
RXTE gain epoch, i.e., with the same high voltage setting of the PCUs
\citep{Jahoda06}.

\subsection{Fourier timing analysis and fitting models.}\label{sec:fourier}
For the Fourier timing analysis we used either the Event modes
E\_125us\_64M\_0\_1s or E\_16us\_64M\_0\_8s, or the Good Xenon data.
Leahy-normalized power spectra were constructed using data segments of
128 seconds and 1/8192~s time bins such that the lowest available
frequency is $1/128 \approx 8 \times 10^{-3}$~Hz and the Nyquist
frequency 4096~Hz.  No background or deadtime corrections were
performed prior to the calculation of the power spectra.  We first
averaged the power spectra \textit{per observation}.  We inspected the
shape of the average power spectra at high frequency ($>2000$~Hz) for
unusual features in addition to the usual Poisson noise. None were
found.  We then subtracted a Poisson noise spectrum estimated from the
power between 3000 and 4000~Hz, using the method developed by
\citet{Kleinwolt04} based on the analytical function of
\citet{Zhang95}. In this frequency range, neither intrinsic noise nor
QPOs are expected based on what we observe in other sources.  The
resulting power spectra were converted to squared fractional rms
\citep{Vanderklis95b}.  In this normalization the power at each
Fourier frequency is an estimate of power density such that the square
root of the integrated power density equals the fractional rms
amplitude of the intrinsic variability in the source count rate in the
frequency range integrated over.

In order to study the behavior of the low-frequency components usually
found in the power spectra of neutron star LMXBs, we needed to improve
the statistics. We therefore averaged observations which were close in
time and had both similar colors and power spectra \citep[see e.g.][
and references within -- see also Appendix in \citealt{Altamirano08}
for a discussion on other possible
methods]{Straaten02,Straaten03,Straaten05,Altamirano05,Altamirano08}.
%

To fit the power spectra, we used a multi-Lorentzian function: the sum
of several Lorentzian components plus, if necessary, a power law to
fit the very low frequency noise (VLFN - see \citealt{Vanderklis06}
for a review).  Each Lorentzian component is denoted as $L_i$, where
$i$ determines the type of component. The characteristic frequency
($\nu_{max}$ as defined below) of $L_i$ is denoted $\nu_i$.  For
example, $L_u$ identifies the upper kHz QPO and $\nu_u$ its
characteristic frequency.  By analogy, other components go by names
such as $L_{\ell}$ (lower kHz), $L_{\ell ow}$ (lower Lorentzian),
$L_{hHz}$ (hectohertz), $L_h$ (hump), $L_b$ (break frequency),
$L_{b2}$ (second break frequency) and their frequencies as
$\nu_{\ell}$, $\nu_{\ell ow}$, $\nu_{hHz}$, $\nu_h$ and $\nu_b$,
$\nu_{b2}$, respectively.  Using this multi-Lorentzian function makes
it straightforward to directly compare the characteristics of the
different components observed in Terzan~2 to those in previous works
which used the same fit function \citep[e.g.,][and references
  therein]{Belloni02,Straaten03,Straaten05,Altamirano05,Altamirano08}.

Unless stated explicitly, we only include those Lorentzians in the
fits whose single trial significance exceeds $3\sigma$ based on the
error in the power integrated from 0 to $\infty$.
We give the frequency of the Lorentzians in terms of characteristic
frequency $\nu_{max}$ as introduced by \citet{Belloni02}: $\nu_{max} =
\sqrt{\nu^2_0 + (FWHM/2)^2} = \nu_0 \sqrt{1 + 1/(4Q^2)}$.  For the
quality factor $Q$ we use the standard definition $Q =
\nu_0/FWHM$. FWHM is the full width at half maximum and $\nu_0$ the
centroid frequency of the Lorentzian. The quoted errors use
$\Delta\chi^2 = 1.0$.  The upper limits quoted in this paper
correspond to a 95\% confidence level ($\Delta\chi^2=2.7$).

\subsection{Energy spectra}

Since the energy spectra of the quiet state (see Section~\ref{sec:lc}) 
of Terzan~2 have already been studied in previous works \citep[see
e.g.][]{Olive98,Barret99,Barret00}, in this paper we concentrate on
the 14 observations that sample the flaring period (see
Section~\ref{sec:lc}). In all 14 cases, we used data of both the PCA
and the HEXTE instruments.

For the PCA, we only used the Standard 2 data of PCU 2, which
was active in all observations. The background was estimated using the
PCABACKEST version 6.0 (see FTOOLS). We calculated the PCU 2
response matrix for each observation using the FTOOLS routine PCARSP V10.1.
For the HEXTE instrument, spectra were accumulated for each cluster
separately.  Dead time corrections of both source and background
spectra were performed using HXTDEAD V6.0. The response matrices were
created using HXTRSP V3.1. 
For both PCA and HEXTE, we filtered out data recorded during, and
up to 30 minutes after  passage through the South
Atlantic Anomaly (SAA). We only use data when the pointing offset from
the source was less than 0.02 degrees and the elevation of the source
respect to the Earth was greater than 10 degrees.  We did not perform
any energy selection prior to the extraction of the spectra. Finally,
we fitted the energy spectra using XSPEC V11.3.2i.

\begin{figure}[!hbtp] 
\center
\resizebox{1\columnwidth}{!}{\rotatebox{0}{\includegraphics{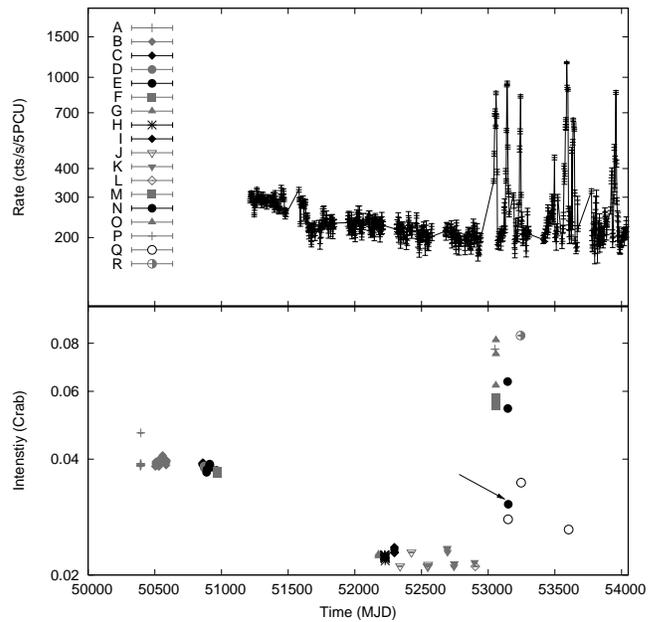}}}
\caption{ \textit{Above:} PCA count rate obtained from the monitoring
observations of Terzan~2 \citep[Monitoring observation of the galactic
bulge - see][]{Swank01}. These data was used for the study of the
long-term variability (see
Sections~\ref{sec:lsp}~\&~\ref{sec:resultslsp}). \textit{Below:}
Terzan~2 intensity (Crab normalized - see
Section~\ref{sec:dataanalysis}) versus time of all pointed
observations. These observations were used for the study of the
0.1--1200~Hz X-ray variability (see Sections~\ref{sec:fourier},
\ref{sec:gralpds}~\&~\ref{sec:kHz}) except for the observation
90058-06-04-00, which is marked with an arrow (see
Section~\ref{sec:results}). The modified Julian date is defined as
$MJD=$ Julian Date$ - 2400000.5$.}
\label{fig:lc}
\end{figure}

\begin{table*}[!hbtp]
\center
\begin{tabular}{|c|c|c|c|c|}\hline 
\hline
 Time interval  & Duration  & Maximum Count Rate   & Flare & N$^o$. of pointed obs.\\
  (MJD)         & (days)      &  (cts/s/5PCU)         & label & sampling the flare\\
\hline
$\sim53038-53071$ & 33 & $\sim850$  & F1& 7\\
$\sim53127-53155$ & 28 & $\sim950$  & F2& 4 \\
$\sim53233-53250$ & 17 & $\sim830$  & F3& 2\\
$\sim53493-53502$ & 9 & $\sim450$   & F4& 0\\
$\sim53566-53602$ & 36 & $\sim1150$ & F5& 1\\
$\sim53631-53651$ & 20 & $\sim650$  & F6& 0\\
$\sim53934-53972$ & 38 & $\sim865$  & F7& 0\\
\hline
\end{tabular}
\caption{Data on the 6 flares observed until MJD~53667. See
Section~\ref{sec:results} for details. (The modified Julian date is
defined as $MJD=$ Julian Date$ - 2400000.5$).}
\label{table:flares}
\end{table*}

\subsection{Search for quasi-periodic variations over days and months}\label{sec:lsp}

Recently, \citet{Wen06} have performed a systematic search for
periodicities in the light curves of 458 sources using data from the
RXTE All Sky Monitor (ASM). Terzan~2 was not included in their
analysis, probably due to the fact that the ASM
source average count rate is low: $2.05\pm0.01$ count/s (the average
of the errors -- $1/n \sum err_i$ -- is 0.8 count/s). 

Since the PCA galactic bulge monitoring \citep{Swank01} has observed
the source for more than 8 years, and the lowest detected source count
rate was $170\pm5$ counts/sec, this new data set provides useful
information to search for long term modulations.
 Lomb-Scargle periodograms \citep{Lomb76,Scargle82,Numerical-Recipes}
as well as the phase dispersion minimization technique \citep[PDM -
see][]{Stellingwerf78} were used. The Lomb-Scargle technique is
ideally suited to look for sinusoidal signals in unevenly sampled
data.  The phase dispersion minimization technique is well suited to
the case of non-sinusoidal time variation covered by irregularly
spaced observations.

\section{Results}\label{sec:results}

\subsection{The light curve}\label{sec:lc}

Figure~\ref{fig:lc} shows both the PCA monitoring lightcurve of the
source \citep[see upper panel and][]{Swank01} and the Crab normalized
intensity (see lower panel and Section~\ref{sec:lcandccd}) of each
pointed observation versus time (in units of modified Julian date $MJD=$
Julian Date$ - 2400000.5$).
In the rest of this paper, we will refer as ``quiet period'' to that
between MJD 51214 and 52945, and as ``flaring period'' between MJD
52945 and 53666.

\begin{figure}[!hbtp] 
\center
\resizebox{1\columnwidth}{!}{\rotatebox{0}{\includegraphics{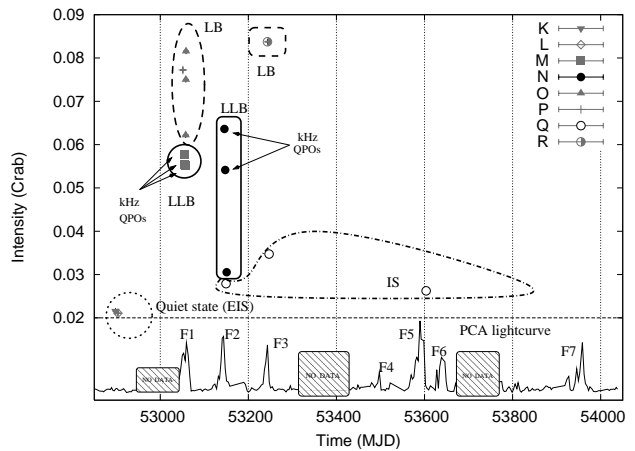}}}
\caption{ \textit{Above:} Intensity in units of the Crab Nebula of the
  observations which were performed during the flare state. The
  different states are labeled (see Section~\ref{sec:state1}). Arrows
  mark the observations in which kHz QPOs were detected (see also
  Section~\ref{sec:kHz} and Table~\ref{table:kHz}).
\textit{Below:} Intensity versus time
showing the seven flares. The data are the same as those of
Figure~\ref{fig:lc}. }
\label{fig:lczoom}
\end{figure}

During the quiet period, 333 monitoring measurements of the source
intensity and 85 pointed observations sample the behavior of the
source. The count rate slowly decreases from an average of $\sim300$,
to an average of $\sim190$ counts$/$s$/$5PCU at an average rate of
$-0.059\pm0.002$ count/s/day.
In the flaring period, 7 flares sampled with 201 monitoring
observations were detected with the galactic bulge scan. 14 pointed
observations partially sampled parts of 4 of these flares.  In
Table~\ref{table:flares} we list approximate dates at which the flux
transitions occurred, the flare durations and the maximum count rates
detected with the PCA.  As mentioned in Section~\ref{sec:lsp}, the
monitoring is done approximately once every three days; additional
gaps in the data are present due to visibility windows.  As of course
we do not have details of flares that may have occurred during these
gaps, the information in Table~\ref{table:flares} is only approximate.
In Figure~\ref{fig:lczoom} we show the intensity of the source during
the flaring period. We label the different flares F1, F2, F3, F4, F5,
F6 and F7 in order of time of occurrence.

We detected three Type I X-ray bursts. One was during the quiet-state
 observation 10090-01-01-021 and two during the flaring-state
 observations 80138-06-06-00 and 90058-06-02-00. A detailed study of
 these X-ray bursts as well as a comparison to bursts observed in
 other sources can be found in \citet{Galloway06}.

\subsection{Color diagrams; identification of states}\label{sec:state1}
\begin{figure}[!hbtp] 
\center
\resizebox{1.0\columnwidth}{!}{\rotatebox{0}{\includegraphics{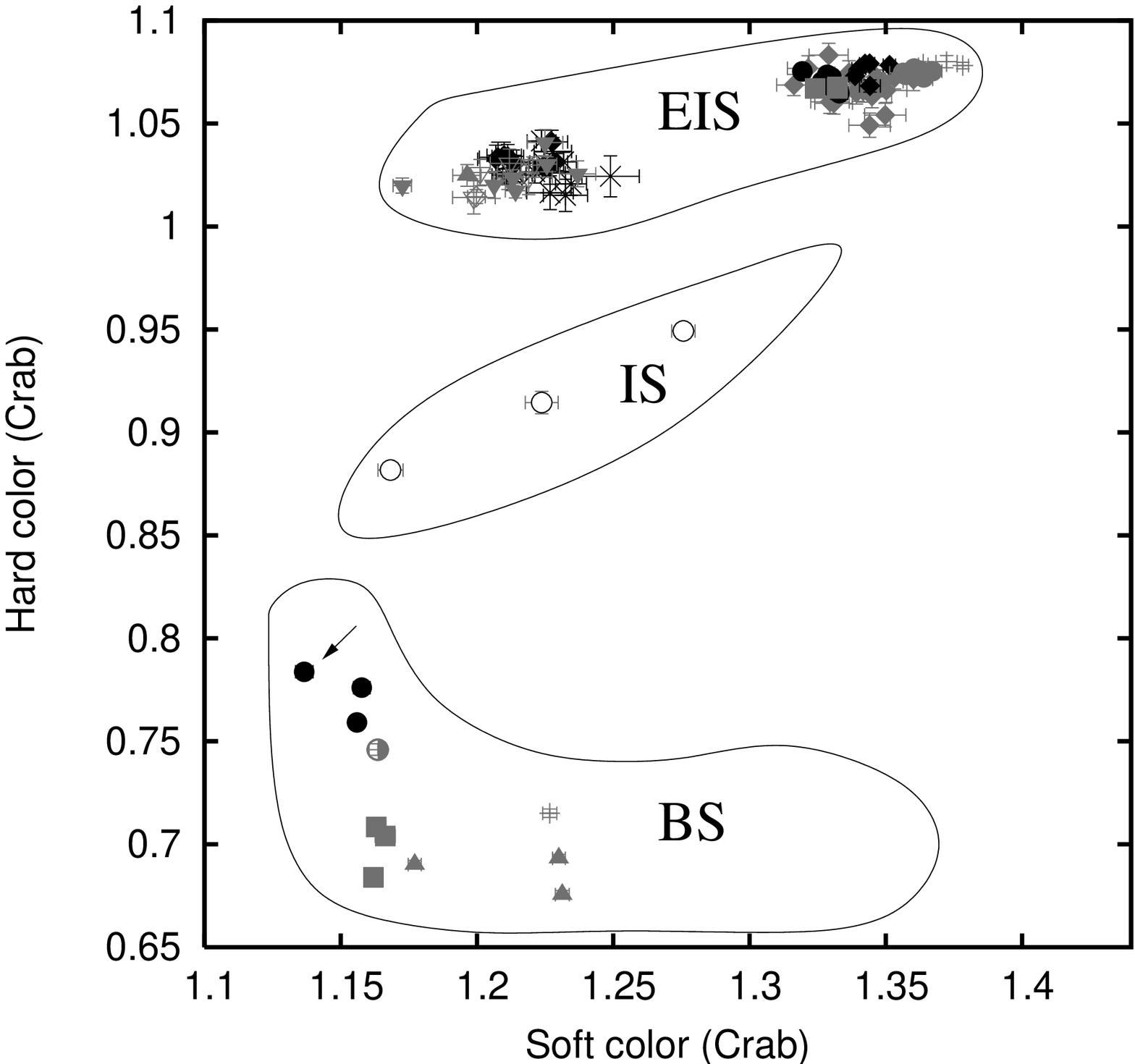}}}
\resizebox{0.7\columnwidth}{!}{\rotatebox{0}{\includegraphics[clip]{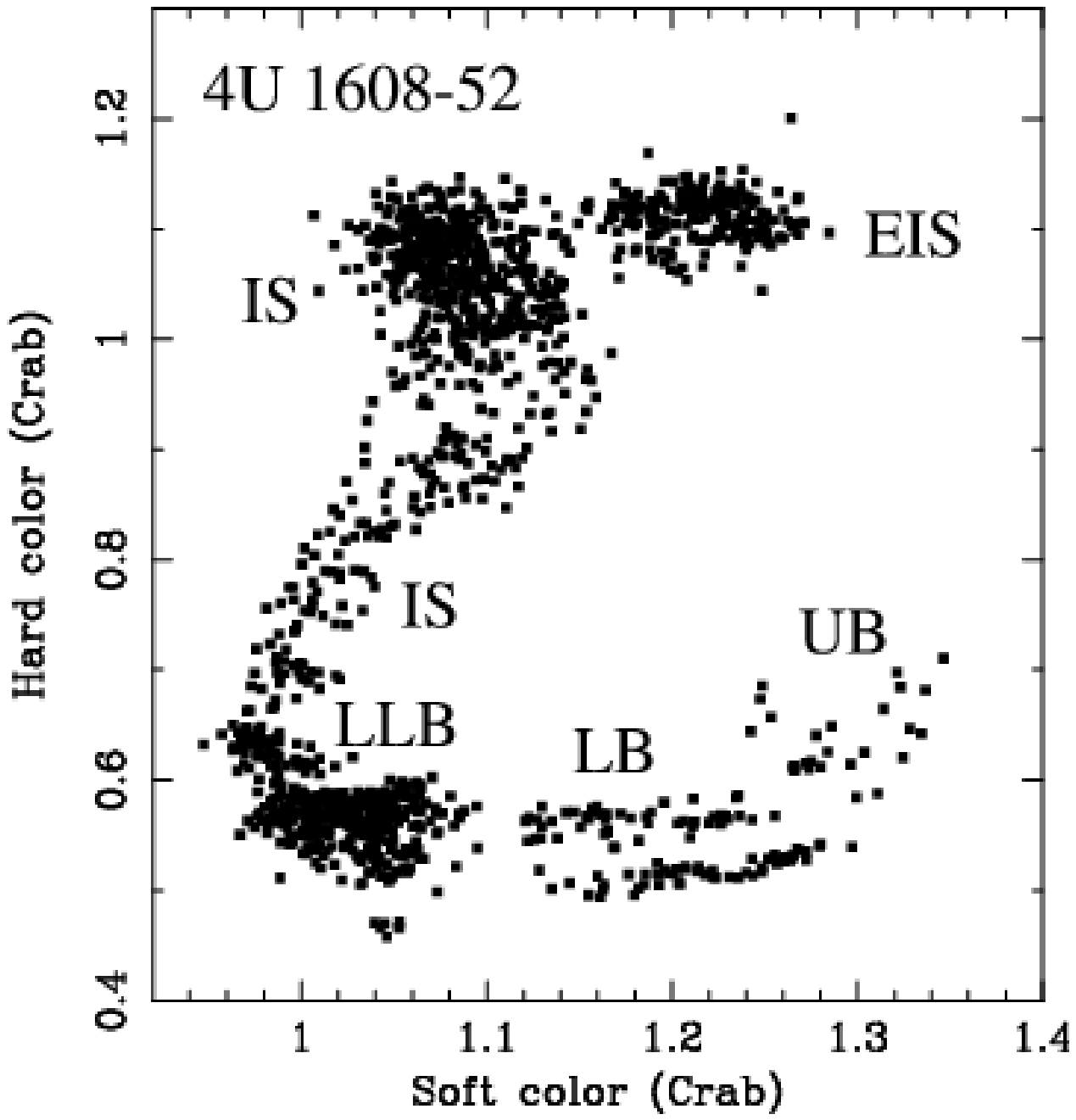}}}
\caption{\textit{Top:} Hard color versus soft color normalized to Crab
as explained in Section~\ref{sec:dataanalysis}.  Different symbols
represent the selections used for averaging the power spectra as
explained in Section~\ref{sec:dataanalysis} and shown in
Figure~\ref{fig:pds} (See Figure~\ref{fig:lc} for symbols). The arrow
marks observation 90058-06-04-00, which was excluded from interval N
(see Section~\ref{sec:results}). \textit{Bottom:} Hard color versus
soft color normalized to Crab for the NS source 4U~1608--52. This
source has been observed in all expected atoll states: extreme island
state (EIS), island state (IS), lower left banana (LLB), lower banana
(LB) and upper banana (UP). The similarity between both figures
suggest that Terzan~2 has been observed in similar states as
4U~1608--52.  }
\label{fig:ccd}
\end{figure}

Figure~\ref{fig:ccd} (top) shows the color-color diagram of the 99
pointed observations. For comparison, we also include the color-color
diagram of the atoll source 4U~1608--52, which
has been observed in all extreme island, island and banana states
(Figure~\ref{fig:ccd}, bottom). The similarity in shape suggests that
Terzan~2 underwent state transitions during observations of the
flares. Based on Figure~\ref{fig:ccd} we can identify the probable
extreme island, island and banana state with the hardest, intermediate
and softest colors, respectively. Since partially sampled patterns in
the color color diagrams are not necessarily unambiguous, \citep[see
review by][]{Vanderklis06}, power spectral analysis (below) is
required to confirm these identifications. The extreme island state
is sampled with 85 pointed observations which are clumped in 2 regions
at similar hard colors but at significantly different soft colors.  We
find 3 observations in the island state. They sampled the lowest
luminosity sections of flares F2, F3 and F5 (see
Figure~\ref{fig:lczoom}).  The banana state is sampled by 11 pointed
observations: 7 during F1, 3 during F2 and 1 during F3.
The identifications above are strengthened by the similarities in
power spectral shapes between Terzan~2 and those reported in other
sources
\citep{Straaten03,Disalvo01,Straaten02,Disalvo03,Straaten05,Linares05,Altamirano05,Migliari05,Altamirano08}.
In the following sections we describe the power spectra in more
detail.

\subsection{kHz QPOs}\label{sec:kHz}

We searched each averaged observation's power spectrum for the
presence of significant kHz QPOs at frequencies $\gtrsim400$~Hz.
As reported in other works \citep{Barret99,Belloni02},
during the quiet period the power spectra of single averaged
observations show significant power up to $\sim300$~Hz which is fitted
with broad ($Q\lesssim0.5$) Lorentzians plus if necessary, one sharp 
Lorentzian to account for $L_{LF}$.
During the flaring period, we found that several observations show
power excess above 400~Hz. For each observation we fitted the averaged
power spectra between 400 and 2000~Hz with a model consisting of one
Lorentzian and a constant to take into account the QPO and Poisson
noise, respectively.
In 5 of the 14 observations that sample the flaring period, we detect
significant QPOs, with single trial significances up to
$6.5\sigma$. In Table~\ref{table:kHz} we present the results of our
fits and information on these observations.  As can be seen, the first
three observations were performed during the rise of the first flare
while the fourth and fifth observations where done during the decay of
the second flare (see Figure~\ref{fig:lczoom}).
As can be seen in Table~\ref{table:kHz}, there are significant
frequency variations. Unfortunately, due to the sparse coverage of the
flares and the fact that we cannot detect the QPO on shorter time
scales than an observation, no further conclusions are possible.

\begin{table*}[!hbtp]
\centering

\begin{tabular}{|c|c|c|c|c|c|c|c|c|}\hline 
\hline
 MJD & ObsId & Used in  & $\nu_0$ (Hz) & FWHM & Q & rms (\%) & Sig.         & c/s \\
     &       & interval &              &      &   &          &              & (5PCU) \\
\hline
53054.94 & 80138-06-02-00 & M & $731\pm3$ & $33.2\pm7.3$&$ 22.1\pm  4.8$ & $10.4\pm0.8$ & $6.5\sigma$  & $\sim665$\\
53055.00 & 80138-06-02-01 & M & $764\pm2$ & $14.3\pm5.2$&$ 53.4\pm 19.4$ & $6.9\pm0.8$  & $4.3\sigma$  & $\sim690$\\
53056.69 & 80138-06-03-00 & M & $772\pm1$ & $15.1\pm4.2$&$ 51.1\pm 14.2$ & $7.1\pm0.6$  & $5.6\sigma$  & $\sim660$\\
53145.57 & 90058-06-01-00 & N & $559\pm2$ & $17.3\pm6.9$&$ 32.3\pm 12.8$ & $6.6\pm0.7$  & $4.7\sigma$  & $\sim765$\\
53147.21 & 90058-06-02-00 & N & $599\pm1$ & $20.3\pm4.8$&$ 29.5\pm  6.9$ & $7.4\pm0.5$  & $6.6\sigma$  & $\sim610$\\
\hline
\end{tabular}
\label{table:kHz}
\caption{Fit results for the 5 averaged observations showing significant
kHz QPOs. }

\end{table*}
%

\begin{figure}[!hbtp] 
\center
\resizebox{1\columnwidth}{!}{\rotatebox{0}{\includegraphics{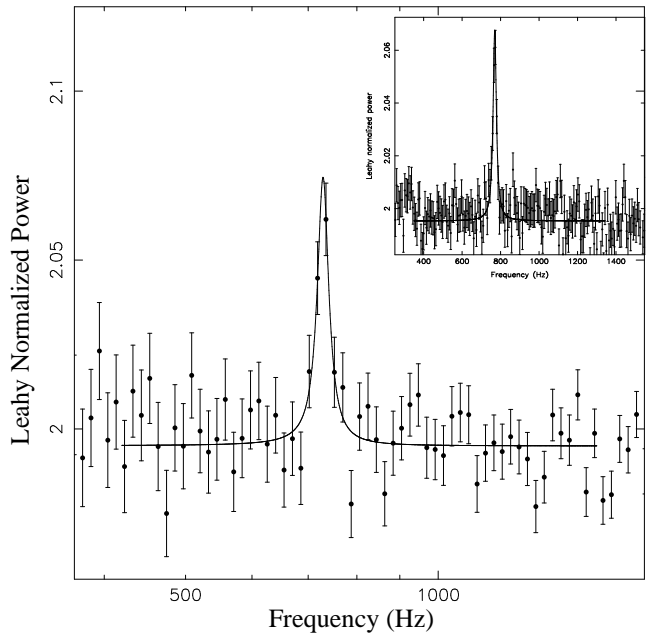}}}
\caption{Fit  to the  kHz QPO  in observation  80138-06-02-00.  In the
subplot on  the top-right of the figure,  we show the fit  to the data
after using the shift-and-add method on all 5 observations reported in
Table~\ref{table:kHz} where the main peak was set to the arbitrary
frequency  $\nu=770$~Hz  (see  Section~\ref{sec:kHz}). No  significant
detections of a second kHz QPO were found. }
\label{fig:kHz}
\end{figure}

We do not significantly detect two simultaneous kHz QPOs in any of the
5 observations.  In order to search for a possible second kHz QPO, we
used the shift--and--add method as described by \citet{Mendez98a}.  We
first tried to trace the detected kilohertz QPO using a dynamical
power spectrum \citep[e.g. see figure 2 in ][]{Berger96} to visualize
the time evolution of the QPO frequency, but the signal was too weak
to be detected on timescales shorter than the averaged observation.
Therefore, for each observation we used the fitted
averaged frequency (see Table~\ref{table:kHz})
to shift each kilohertz QPO to the arbitrary frequency of 770~Hz.
Next, the shifted, aligned, power spectra were averaged. The average
power spectrum was finally fitted in the range 300--2048~Hz so as to
exclude the edges, which are distorted due to the shifting method.  To
fit the averaged power spectrum, we used a function consisting of a
Lorentzian and a constant to fit the QPO and the Poisson noise,
respectively.  We studied the residuals of the fit, but no significant
power excess was present apart from the 770~Hz feature. In
Figure~\ref{fig:kHz} we show the fitted kHz QPO for observation
80138-06-02-00 (no shift and add was applied) and the
shifted-and-added kHz QPO detected with the method mentioned above.
%
Since it is known that the kHz QPOs become stronger at higher energies
\citep[e.g.][and references within]{Berger96,Mendez98,Vanderklis00},
we repeated the analysis described above (which was performed on the
full PCA energy range), using only data at energies higher than
$\sim6$~keV or higher than $\sim10$~keV.  Again, no significant second
QPO was present.
It is important to note that this method can produce ambiguous results as we
cannot be sure we are always shifting  the same component (either $L_u$
or $L_{\ell}$).  We also tried different subgroups, i.e. adding only
two to three different observations, but found the same results.

\begin{figure}[!hbtp] 
\center
\resizebox{1\columnwidth}{!}{\rotatebox{-90}{\includegraphics{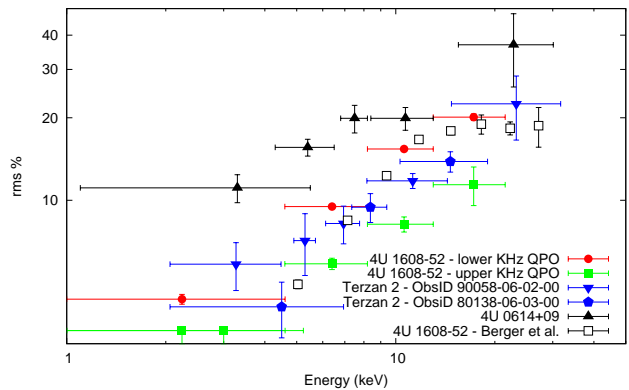}}}
\caption{Energy dependence of two representative kHz QPOs of
Terzan~2. One in flare F1 (pentagons -- OBSid 80138-06-03-00) and
another in flare F2 (downward pointed triangles -- OBSid 90058-06-02-00). For
comparison we show the energy dependence of both lower (circles \&
open squares) and upper (squares) peak of the atoll source
4U~1608--52 \citep{Berger96,Mendez98a} and the upper peak (upwards pointed
triangles) of the atoll source 4U~0614+09
\citep{Mendez97,Straaten00}. }
\label{fig:rmsvse}
\end{figure}

To investigate the energy dependence of the kHz QPOs, we
divided each power spectrum into 3, 4 or 5 energy intervals in order
to have approximately the same count rate in all the intervals.  We
then produced the power spectrum as described in
Section~\ref{sec:dataanalysis} and refitted the data where both
frequency and  Q were fixed to the values obtained for
the full energy range  (see Table~\ref{table:kHz}).  In
Figure~\ref{fig:rmsvse} we show the results for the representative kHz
QPOs in flares F1 and F2 (observations 80138-06-03-00 and
90058-06-02-00, respectively). 
Similarly to what is observed in other sources, the fractional rms
amplitude of the kHz QPOs increases with energy.
The data show that there is no significant
difference in the energy dependence of the kHz QPO at $\sim599$~Hz
with that at $\sim772$~Hz.
%
%

\subsection{Averaged power spectrum}\label{sec:gralpds}

\begin{figure*}[!hbtp] 
\center
\resizebox{1.4\columnwidth}{!}{\rotatebox{0}{\includegraphics[clip]{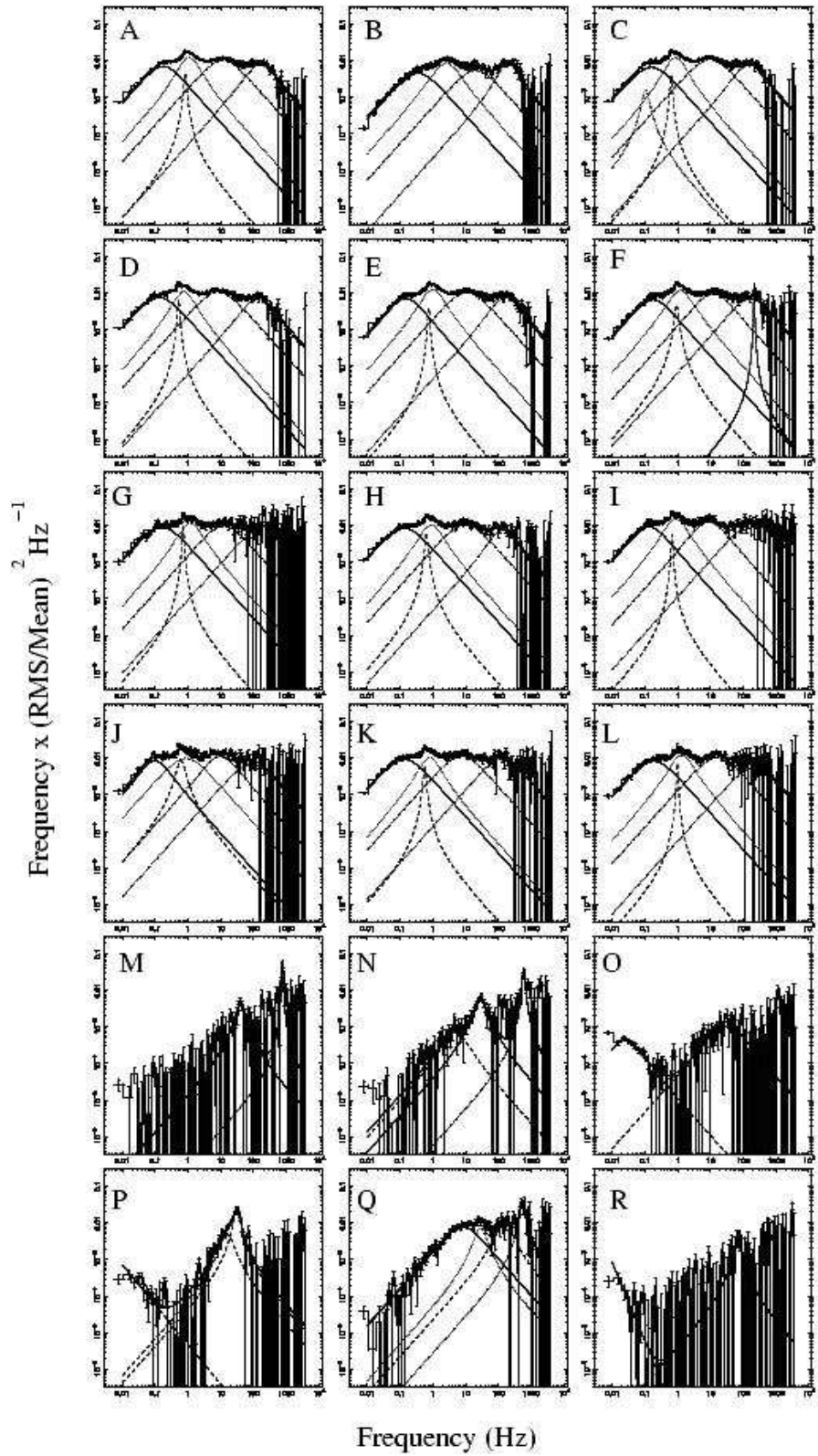}}}
\caption{Power spectra and fit functions in the power spectral density
  times frequency representation.  Each plot corresponds to a
  different region in the color-color and color-intensity diagrams
  (see Figures~\ref{fig:ccd} and~\ref{fig:lc}).  The curves mark the
  individual Lorentzian components of the fit.  The re-binning varies
  between the panels to balance the frequency resolution with the
  signal-to-noise ratio. For a detailed identification, see
  Table~\ref{table:data}. }
\label{fig:pds}
\end{figure*}

As mentioned in Section~\ref{sec:fourier}, in order to study the
behavior of the low-frequency components usually found in the power
spectra of neutron star LMXBs, we needed to improve the statistics. We
therefore averaged observations which were close in time and had both
similar colors and power spectra.
%
%
The resulting data selections are labeled from A to R (ordered mainly
in time -- see Table~\ref{table:averages} for details on which
observations were used for each interval and their colors).  Their
corresponding average power spectra are displayed in
Figure~\ref{fig:pds}.

\begin{table*}[!hbtp]
\center
{
\begin{tabular}{|c|c|c|c|}\hline 
\multicolumn{4}{|c|}{Interval A} \\
\hline
 Observation  & Soft color (Crab) & Hard color (Crab) & Intensity (Crab) \\
\hline
10090-01-01-000 & $ 1.3615\pm0.0017$ & $ 1.0760\pm0.0013$ & $ 0.0389\pm0.0001$ \\  
10090-01-01-001 & $ 1.3508\pm0.0019$ & $ 1.0761\pm0.0014$ & $ 0.0384\pm0.0001$ \\  
10090-01-01-00 & $ 1.3619\pm0.0019$ & $ 1.0757\pm0.0015$ & $ 0.0384\pm0.0001$ \\  
10090-01-01-020 & $ 1.3783\pm0.0021$ & $ 1.0781\pm0.0016$ & $ 0.0386\pm0.0001$ \\  
10090-01-01-021 & $ 1.4435\pm0.0019$ & $ 0.9846\pm0.0013$ & $ 0.0468\pm0.0001$ \\  
10090-01-01-022 & $ 1.3638\pm0.0017$ & $ 1.0788\pm0.0013$ & $ 0.0387\pm0.0001$ \\  
10090-01-01-02 & $ 1.3723\pm0.0039$ & $ 1.0801\pm0.0029$ & $ 0.0389\pm0.0001$ \\  
\hline
\multicolumn{4}{|c|}{Interval B} \\
\hline
20170-05-01-00 & $ 1.3501\pm0.0075$ & $ 1.0660\pm0.0058$ & $ 0.0383\pm0.0001$ \\  
20170-05-02-00 & $ 1.3440\pm0.0077$ & $ 1.0492\pm0.0059$ & $ 0.0384\pm0.0001$ \\  
20170-05-03-00 & $ 1.3498\pm0.0076$ & $ 1.0541\pm0.0058$ & $ 0.0389\pm0.0001$ \\  
20170-05-04-00 & $ 1.3449\pm0.0073$ & $ 1.0632\pm0.0056$ & $ 0.0394\pm0.0001$ \\  
20170-05-05-00 & $ 1.3492\pm0.0071$ & $ 1.0703\pm0.0055$ & $ 0.0398\pm0.0001$ \\  
20170-05-06-00 & $ 1.3392\pm0.0072$ & $ 1.0650\pm0.0056$ & $ 0.0394\pm0.0001$ \\  
20170-05-07-00 & $ 1.3465\pm0.0072$ & $ 1.0726\pm0.0055$ & $ 0.0393\pm0.0001$ \\  
20170-05-08-00 & $ 1.3295\pm0.0071$ & $ 1.0605\pm0.0056$ & $ 0.0382\pm0.0001$ \\  
20170-05-09-00 & $ 1.3398\pm0.0070$ & $ 1.0704\pm0.0054$ & $ 0.0389\pm0.0001$ \\  
20170-05-10-00 & $ 1.3163\pm0.0064$ & $ 1.0687\pm0.0051$ & $ 0.0387\pm0.0001$ \\  
20170-05-11-00 & $ 1.3601\pm0.0071$ & $ 1.0714\pm0.0054$ & $ 0.0400\pm0.0001$ \\  
20170-05-12-00 & $ 1.3365\pm0.0074$ & $ 1.0747\pm0.0058$ & $ 0.0399\pm0.0001$ \\  
20170-05-13-00 & $ 1.3532\pm0.0070$ & $ 1.0721\pm0.0053$ & $ 0.0409\pm0.0001$ \\  
20170-05-14-00 & $ 1.3215\pm0.0076$ & $ 1.0768\pm0.0061$ & $ 0.0407\pm0.0001$ \\  
20170-05-15-00 & $ 1.3417\pm0.0070$ & $ 1.0658\pm0.0054$ & $ 0.0394\pm0.0001$ \\  
20170-05-16-00 & $ 1.3398\pm0.0081$ & $ 1.0669\pm0.0062$ & $ 0.0398\pm0.0001$ \\  
20170-05-17-00 & $ 1.3290\pm0.0072$ & $ 1.0832\pm0.0057$ & $ 0.0386\pm0.0001$ \\  
20170-05-18-00 & $ 1.3318\pm0.0086$ & $ 1.0682\pm0.0067$ & $ 0.0392\pm0.0001$ \\  
20170-05-19-00 & $ 1.3336\pm0.0068$ & $ 1.0667\pm0.0053$ & $ 0.0395\pm0.0001$ \\  
20170-05-20-00 & $ 1.3309\pm0.0069$ & $ 1.0599\pm0.0054$ & $ 0.0397\pm0.0001$ \\  
\hline
\end{tabular}
\caption{Observations used for the timing analysis. The colors and
intensity are corrected by dead time and normalized to the Crab Nebula
(see Section~\ref{sec:dataanalysis}).  The complete table can
be obtained digitally from ApJ.}
\label{table:averages}
}
\end{table*}

\subsubsection{The power spectra of the quiet period}\label{pdsqp}

Intervals A to L are all part of the quiet segment. 4 to 6
components were needed to fit all the power spectra of this group,
where 11 out of the 12 power spectra showed significant broad components at
$\sim150$, $\sim10$, $\sim1$ and at $\sim0.2$~Hz. 
 Interval G is the
exception where the broad component at $\sim150$~Hz was not
significantly detected ($8.5$\% rms-amplitude upper limit).  A QPO
($Q\gtrsim2$) at $\sim0.8$~Hz was also significantly
detected in 11 out of 12 power spectra, interval B being the exception ($1.4$\%
rms-amplitude upper limit).
%
Finally, an extra component ($L_{vl}$, where vl stands for ``very
low'' ) at frequency $\nu_{vl}\sim0.1$~Hz was detected only in
interval C.

\begin{figure}[!hbtp] 
\center
\resizebox{1\columnwidth}{!}{\rotatebox{0}{\includegraphics[clip]{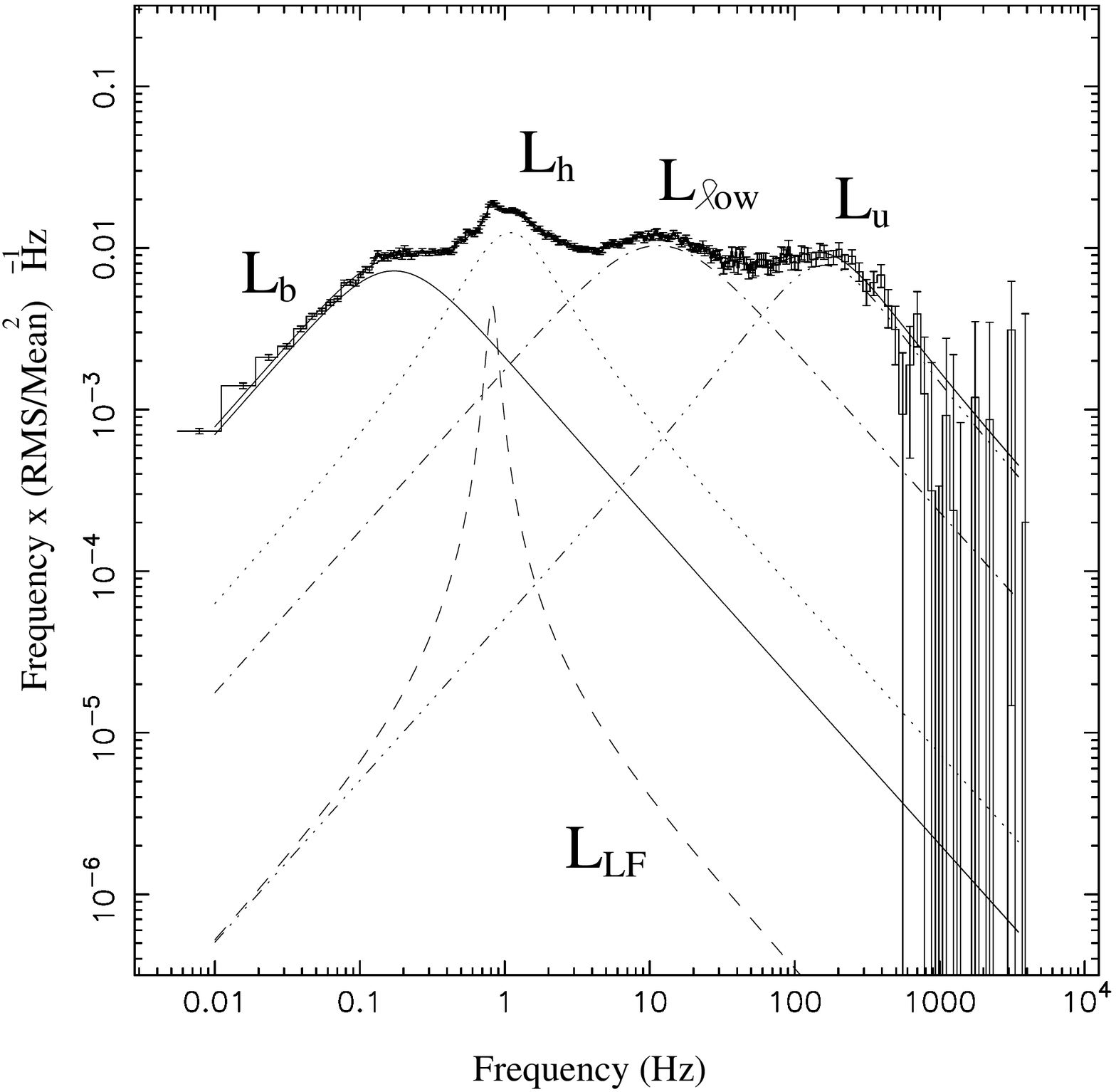}}}
\caption{Example of the typical power spectra of the quiet period. The
  main 5 Lorentzians are $L_u$, $L_{\ell ow}$, $L_h$, $L_{LF}$ and
  $L_b$ (see Section~\ref{pdsqp}).}
\label{fig:syntetic}
\end{figure}

\begin{figure}[!hbtp] 
\center
\resizebox{1\columnwidth}{!}{\rotatebox{0}{\includegraphics{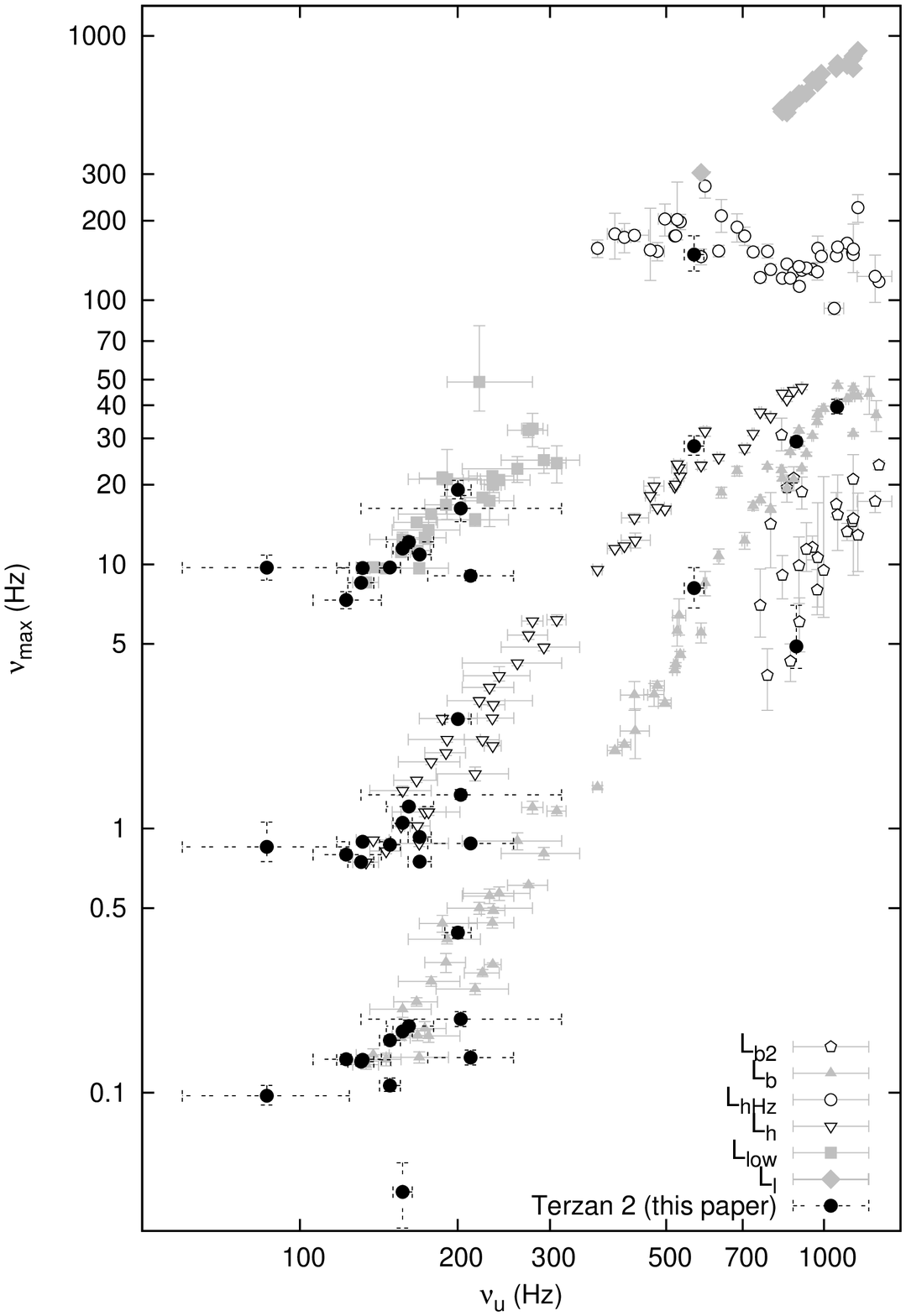}}}
\caption{The characteristic frequencies $\nu_{max}$ of the various
  power spectral components plotted versus $\nu_{u}$. The grey symbols
  mark the atoll sources 4U~0614+09, 4U~1728--34 \citep{Straaten02},
  4U~1608--52 \citep{Straaten03}, Aql X-1 \citep{Reig04} and
  4U~1636--53 \citep{Altamirano08} and the low luminosity bursters
  Terzan~2 (previous results), GS~1826--24 and SLX~1735--269
  \citep[][but also see \citealt{Belloni02}]{Straaten05}.  The black
  bullets mark our results for Terzan~2. Note that we only plot the
  results for Intervals A-N and Q (note that as mentioned in
  Section~\ref{sec:tfp}, the points for intervals M and N
  ($\nu_u>800$~Hz) were plotted under the assumption that $\nu_u =
  \nu_{\ell} + 300$~Hz). For intervals O, P and R no kHz QPOs were
  detected (see Section~\ref{sec:gralpds}). }
\label{fig:nuvsnu}
\end{figure}

Similar power spectra have been reported in the extreme island state of
the atoll sources 4U~0614+09, 4U~1728--34 \citep{Straaten02},
XTE~J1118+480, SLX~1735--269 \citep{Belloni02} and 4U~1608--52
\citep{Straaten03}. \citet{Olive98} and \citet{Belloni02} analyzed 
an early subset of the data we present in this work. Our results are consistent
with the frequencies reported in those works.
Following \citet{Straaten02}, \citet{Belloni02} and
\citet{Straaten03}, we identify the components as $L_u$, $L_{\ell
ow}$, $L_h$, $L_{LF}$ and $L_b$, where $\nu_u\sim150$~Hz, $\nu_{\ell
ow}\sim10$~Hz, $\nu_h\sim1$~Hz, $\nu_{LF}\sim0.8$~Hz and
$\nu_b\sim0.1$~Hz, respectively.  These identifications are
strengthened by the correlations shown in Figure~\ref{fig:nuvsnu}, in
which we plot the characteristic frequency of all components versus
that of $\nu_u$ for several atoll sources (see Section~\ref{sec:tfp});
clearly, we observe the same components as seen in the EIS in other
atoll sources. 
In Figure~\ref{fig:syntetic} we provide a modeled power spectra were
the main 5 Lorentzians (i.e.  $L_u$, $L_{\ell ow}$, $L_h$, $L_{LF}$
and $L_b$) are represented. Components in Figures~\ref{fig:pds}
and~\ref{fig:syntetic} are represented with the same line style.

\subsubsection{The power spectra of the flaring period}\label{sec:flaringsegment}
Intervals M, N, O, P, Q and R are part of the flaring period.
For intervals M, N and Q, high frequency ($\gtrsim400$~Hz) single QPOs
are seen which can be identified as either  the lower or the upper kHz
QPO.

The kHz QPOs in interval M and N correspond to the averages of
significant QPOs observed in single observations. As seen in
Table~\ref{table:kHz}, the characteristic frequency of the kHz QPOs
averaged in each of the two intervals are within a range of 50~Hz. By
our averaging method we are affecting the Q value of the kHz QPOs but
improving the statistics for measuring the characteristics of the
features at lower frequencies \citep[see Appendix in ][ for a
discussion on this issue]{Altamirano08}.

Interval Q is an average of three single observations (see
Table~\ref{table:averages}) that individually do not show significant
QPOs ( $Q>2$) at high frequencies although low-Q power excess can be
measured. $L_h$ in interval Q is only $2.6\sigma$ significant (single
trial) but required for a stable fit.

 Interval O shows a broad component at $30.5\pm6.3$~Hz and a 3.3\% rms
low-frequency noise. In this case, the very low frequency noise was fitted
with a broad Lorentzian because a power law gave an unstable fit. 
The excess of power at
$\nu\gtrsim1000$~Hz is less than $3\sigma$ significant.
Interval P shows a power spectrum with a power-law  low
frequency noise, and two Lorentzian components at frequencies
$15.5\pm1.6$ and $30.5\pm6.3$~Hz (see Figure~\ref{fig:pds}). In this
case the high $\chi^2/dof = 218 / 163$ reveals that the Lorentzians do
not satisfactorily fit the data.  As can be seen in
Figure~\ref{fig:pds}, there is a steep decay of the power above
$\nu\sim35$~Hz and power excess at $\sim70$~Hz. A fit with three
Lorentzians becomes unstable.  To further investigate this, we
refitted the power spectrum using instead two Gaussians and one power
law to fit the power at $\nu\lesssim40$~Hz, and one Lorentzian to fit
the possible extra component at $\sim70$~Hz.  The steeper Gaussian
function better fits the steep power decay than the Lorentzians. In
this fit, with three more free parameters, we obtain a
$\chi^2/dof=188/160$,  and the
Lorentzian at $69.1\pm2.6$~Hz becomes $3.4\sigma$ single-trial
significant. This power spectrum is very similar to those reported by
\citet{Migliari04,Migliari05} and \citet{Migliari05} for the atoll
sources 4U~1820--30 and Ser X-1, respectively. Besides the similarity
in shape of the power spectrum, it is interesting to note that in both
cases these authors found a best fit with components at similar
frequencies to the ones we observe in Terzan~2 and
which they interpreted as $\nu_b$, $\nu_h$ and $\nu_{hHz}$. This
coincidence suggests that the sources were in very similar states. To
our knowledge, no systematic study of this state has been reported as
yet.

Interval R consists of only one observation (90058-06-05-00) of
$\sim1.4$ ksec of data.  In addition to a power law with index
$\alpha=3.1\pm0.7$, we detect one Lorentzian at $68.5\pm7.2$~Hz. Its
frequency is rather high if we compare it with the other power spectra
presented in this paper and even when compared with results in other
sources (see Figure~\ref{fig:nuvsnu}). This result might be due to 
blending of components due to the low statistics present in this power
spectrum.

From the 99 pointed observations, only observation 90058-06-04-00
was not included in any of the averages described above. The averaged
colors of this observation are similar to those of Interval N
(observations 90058-06-01-00, 90058-06-02-00) but
the power spectrum does not show a significant QPO at $\simeq560$~Hz and
can be fitted with a single Lorentzian at $\nu_{max}=23.6\pm3.3$~Hz,
$Q=0.5\pm0.2$ and rms$=13.2\pm0.9$\%.  The residuals of the fit show
excess power at $\simeq800$~Hz but no significant kHz QPO. 
Although the colors of observation 90058-06-04-00 are similar from
those of interval N, the power spectrum is different. Because of this,
we refrained from averaging 90058-06-04-00 into interval N.

\begin{figure}[!hbtp] 
\center
\resizebox{1\columnwidth}{!}{\rotatebox{-90}{\includegraphics{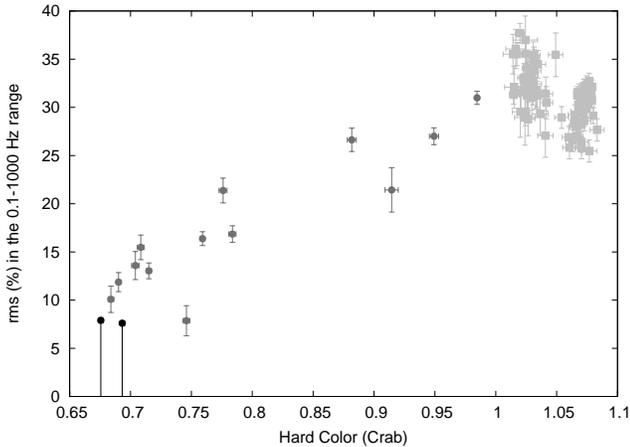}}}
\caption{Fractional rms (\%) amplitude versus hard color normalized to the Crab
Nebula as explained in Section~\ref{sec:dataanalysis}.  The light-grey
squares represent the data of the ``quiet'' state while the dark-grey
circles represent the data for the ``flare'' state.  The black circles
represent the upper limit ($\Delta\chi^2=2.7$) for the observations
80138-06-05-00 and 80138-06-06-00 which also partially sampled the
flare states.}
\label{fig:rms1000}
\end{figure}

\subsection{Integrated power}

In order to study the rms amplitude dependence on color and intensity,
we calculated the average integral power per observation between 0.1
and 1000~Hz.  In Figure~\ref{fig:rms1000} we show the 0.01--1000~Hz
averaged rms amplitude (\%), of each of the 99 observations, versus
its average hard color.  The observations which sample the island and
banana state correlate with the averaged hard color.  This type of
correlation has been already observed in black hole candidates and
neutron star systems \citep[see e.g.][]{Homan01}.  At colors harder
than 1.0, there are two clumps (grey squares in
Figure~\ref{fig:rms1000}) which can also be seen in the color-color
diagram (Figure~\ref{fig:ccd}).  They both correspond to observations
of the extreme island state (quiet state) of the source. As we show in
Figure~\ref{fig:pds}, the power spectral shape remains approximately
the same with time.  However, for a given hard color, the total rms
amplitude can change up to $\simeq30$\%.
No correlation with time, intensity or soft color was found and these
changes are seen within a small range in intensity (less than 4
mCrab). Similar results are also observed when the rms amplitude is
calculated in the 0.01--300~Hz range.

\subsection{Comparing Terzan~2 with other LMXBs}

\subsubsection{The flaring period}\label{sec:tfp}

\begin{figure}[!hbtp] 
\center
\resizebox{1\columnwidth}{!}{\rotatebox{-90}{\includegraphics{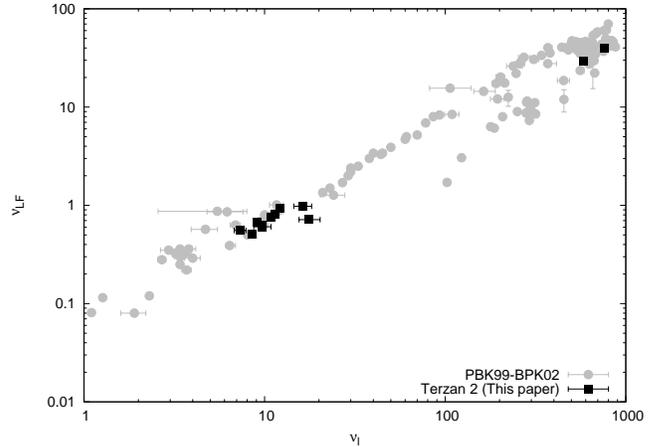}}}
\caption{PBK relation after \citet[PBK99 -][]{Psaltis99} and
\citet[BPK02 -][]{Belloni02}. The black squares at $\nu_{LF}<10$~Hz
represent the data of Intervals A and C--L while the black squares at
$\nu_{LF}>20$~Hz represent the data of Intervals M and N. In Interval
B we do not detect $L_{LF}$ and in intervals Q--R we do not detect
$L_{\ell}$.}
\label{fig:pbk}
\end{figure}

In Figure~\ref{fig:pbk} we plot the frequency correlations between
$L_{LF}$ and $L_{\ell}$ reported by \citet{Psaltis99} and updated by
\citet{Belloni02}. In Figure~\ref{fig:nuvsnu} we plot the
characteristic frequency of all components versus that of $\nu_u$ for
the atoll sources 4U~0614+09, 4U~1728--34, 4U~1608--52, 4U~1636--53
and Aql~X-1 and the two low luminosity bursters GS~1826--24 and
SLX~1735--269 \citep{Straaten02,Straaten03,Reig04,Altamirano08}.
Using these correlations to identify the highest frequencies we
observe (in intervals M, N \& Q) as either $\nu_u$ or $\nu_{\ell}$
presents a problem. Both interpretations give consistent results since
the correlations observed are complex when $\nu_u \gtrsim600$~Hz
\citep{Straaten05}.

In recent work, \citet{Barret05a,Barret05b,Barret05c} have
systematically studied the variation of the frequency, rms amplitude
and the quality factor Q of the lower and upper kHz QPOs in the
low-mass X-ray binaries 4U~1636--536, 4U~1608--52 and
4U~1735--44. Although Q depends on the frequency of the component,
these authors show that $Q_{\ell}$ is always above $\sim30$, while
$Q_u$ is generally below $\sim25$. By comparing our results for the
kHz QPOs with those of \citet{Barret05a,Barret05b,Barret05c}, we find
that the 5 QPOs listed in Table~\ref{table:kHz} (and averaged in
intervals M and N) are relatively high-Q and hence consistent with
being $L_{\ell}$, but also still consistent with being $L_u$. The
quality factor of the kHz QPO found in interval Q is low ($2.4\pm0.6$)
and hence this kHz QPO is probably $L_u$.

\begin{figure*}[!hbtp] 
\center
\resizebox{1.5\columnwidth}{!}{\rotatebox{0}{\includegraphics{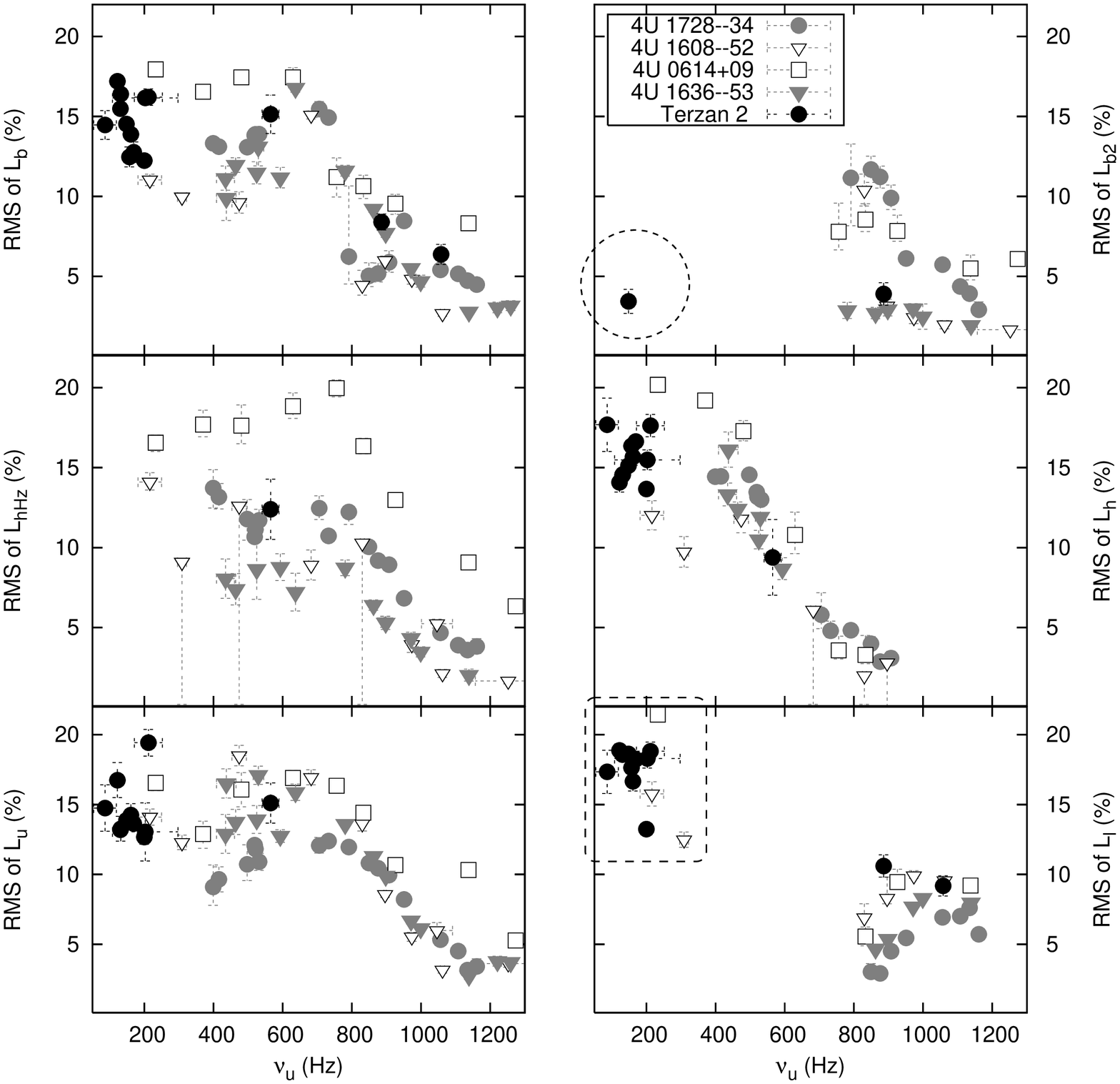}}}
\caption{The fractional rms amplitude of all components (except
$L_{LF}$) plotted versus $\nu_{u}$.  The symbols are labeled in the
plot. The data for 4U~1728--34, 4U~1608--52 and 4U~0614+09 were taken
from \citet{Straaten05}. The data for 4U~1636--53 were taken from
\citet{Altamirano08}. Note that for $L_{hHz}$ and $L_{h}$ of
4U~1608--52, the 3 triangles with vertical error bars which intersect
the abscissa represent 95\% confidence upper limits (see
\citet{Straaten03} for a discussion). The points inside the circle
represent our results for $L_{vl}$ while the points inside the square
represent results for $L_{low}$ (see Section~\ref{sec:discussion}
for a discussion). Note that as mentioned in Section~\ref{sec:tfp},
the points for intervals M and N ($\nu_u>800$~Hz) were plotted under
the assumption that $\nu_u = \nu_{\ell} + 300$~Hz. }
\label{fig:rms}
\end{figure*}

In Figure \ref{fig:rms} we plot the fractional rms amplitude of all
components (except $L_{LF}$) versus $\nu_{u}$ for the 4 atoll sources
4U~0614+09, 4U~1728--34, 4U~1608--52, 4U~1636--53. With respect to the
kHz QPO identification, the interpretation that we found $L_{\ell}$ in
interval N is strengthened by the rms amplitudes of the two
low-frequency components found in the averaged power spectrum. If the
QPO we observe is not $L_{\ell}$ but $L_u$, then the low frequency QPO
pairs can be identified as either $L_h$-$L_b$ or $L_b$-$L_{b2}$. For
$\nu_u=586.1\pm6.6$, the pair $L_h$-$L_b$ is not consistent with what
we observe for other atoll sources (Figure~\ref{fig:rms}), since $L_b$
is not seen with rms amplitude as low as $3.9^{+0.8}_{-0.5}$\%. The
pair $L_b$-$L_{b2}$ is not consistent with the data either, since
$L_{b2}$ is always observed at $\nu_u\gtrsim800$~Hz. If the QPO is
$L_{\ell}$, then based on what we observe in other well-studied
sources
\citep{Mendez98a,Mendez99b,Disalvo03,Barret05b,Vanderklis06,Mendez07}
we expect that the frequency difference between kHz QPOs is
$\Delta_\nu = \nu_u - \nu_{\ell} \simeq 300$ and therefore $\nu_u
\simeq \nu_{\ell} + 300 \simeq 586 + 300 \simeq 886$~Hz.
 Under this assumption (i.e. $\Delta_\nu \simeq 300$) and by using the
 same reasoning as above, then only the pair $L_b$-$L_{b2}$ is
 consistent with the data. In the case of interval M, only one
 component is found at low frequencies with an rms amplitude of
 $6.4\pm0.6$\%. This result is consistent with several interpretations
 when compared with the data shown in Figure~\ref{fig:rms}. Therefore,
 for interval M we cannot improve confidence in the identification of
 the kHz QPO using Figure~\ref{fig:rms}.

In Figures~\ref{fig:nuvsnu}, \ref{fig:pbk} \& \ref{fig:rms} we have
plotted the data for Terzan~2 based on the identifications above. As
discussed later in this paper, such identifications need to be
confirmed.

\subsubsection{The quiet period}

In Figure \ref{fig:rms} we show that the rms amplitude of $L_u$,
$L_{\ell ow}$, $L_h$ and $L_b$ in Terzan~2 approximately follow the
trend observed for other sources. Since \citet{Psaltis99} interprets
$L_{\ell ow}$ and $L_{\ell}$ as the same component in different source
states, in Figure~\ref{fig:rms} we plot the data for both components
together. Of course, in Figure~\ref{fig:rms} and
Figure~\ref{fig:nuvsnu} as well, there is the well-known gap between
these two components.
Regarding lower frequency components, the point inside the circle
represents our result for $L_{vl}$ in interval C, which is the weakest
component found in the EIS of Terzan~2.  We plotted our results with
those for $L_{b2}$. This component might be related with the VLFN
Lorentzian \citep[see][]{Schnerr03,Reerink05}.
\citet{Altamirano08} showed that the rms amplitude of $L_{LF}$ of
several atoll sources does not correlate with $\nu_u$.  There is also
no evidence for a correlation in the case of Terzan~2 (not plotted).

\subsection{Luminosity estimates from spectral fitting}

We fitted the PCA and the two cluster's HEXTE spectra simultaneously
using a model consisting of a black body to account for the soft
component of the spectra and a power law to account for the hard
component.  In some cases, it was necessary to add a Gaussian to take
into account the iron K$\alpha$ line \citep[6.4 keV, see
e.g.][]{White86}.  We ignore energies below 2.5 and above 25keV for
the PCA spectra, and below 20 and above 200 keV for the HEXTE spectra
\citep[see e.g.][]{Barret00}.  In most cases, it is not possible to
well constrain the interstellar absorption $n_H$ if we lack spectral
information below 2.5 keV. We therefore opted to fix $n_H$ to the
value $n_H=1.2\cdot10^{22}$ H atoms cm$^{-2}$ (in the Wisconsin cross
section wabs model -- see \citealt{Morrison83} ) based on previous
ASCA/BeppoSAX results \citep[]{Olive98,Barret99,Barret00}.

Assuming a distance of 6.6~kpc, we found that all 14 observations have
luminosities between $\sim0.4$ and $\sim1.35 \times 10^{37}$ erg
s$^{-1}$ in the energy range 2--20 keV. We also found that at high
energies (20--200 keV) the luminosities of most observations were less
than $0.09 \times 10^{37}$ erg s$^{-1}$. The exceptions are the three
observations which sample the island state, which show 20--200~keV
luminosities of $\sim0.16$, $\sim0.23$ and $\sim0.29 \times 10^{37}$
erg s$^{-1}$ (observations 90058-06-03-00, 90058-06-06-00 and
91050-07-01-00, respectively).
This may be compared with observations of the brightest interval of
 the quiet period of Terzan~2, which have averaged luminosities
 $L_{1-20 keV} = 0.81\times10^{37}$ erg s$^{-1}$ and $L_{20-200 keV} =
 0.48 \times10^{37}$ erg s$^{-1}$ \citep{Barret00}. Clearly, in
 between the flares the luminosity can drop to similarly low values as
 in the quiet period. This is consistent with the lightcurve we show
 in Figure~\ref{fig:lc}. We note that the observations studied by
 \citet{Barret00} correspond to MJDs $50391-50395$ (November 4-8,
 1996) and sample the brightest part of the quiet period of this
 source observed with RXTE (see Figure~\ref{fig:lc}).

We are particularly interested in the luminosity at which the kHz QPOs
are detected in Terzan~2 compared with other sources. \citet{Ford00}
have measured simultaneously the properties of the energy spectra and
the frequencies of the kHz QPOs in 15 low-mass X-ray binaries covering
a wide range of X-ray luminosities. The observations of intervals M
and N (see Table~\ref{table:flares}) have average luminosities
$L_{2-50 keV} / L_{edd}$ between 0.025 and 0.04 (where
$L_{edd}=2.5\times10^{38}$ erg s$^{-1}$).  The three observations
that sample the island state (interval Q) have average luminosity
$L_{2-50 keV} / L_{edd} \simeq 0.02$. This means that during the flares,
Terzan~2 shows kHz QPOs at similar luminosities to the atoll sources
Aql X-1, 4U~1608--52, 4U~1702--42 and 4U~1728--34 \citep[see figure 1
in ][]{Ford00}. 

\subsection{Lomb Scargle Periodograms}\label{sec:resultslsp}

During the quiet period (51214--52945 MJD), we found no significant
periodicities using either  the Lomb-Scargle or the PDM techniques in
the full data set nor in sub-intervals.

\begin{figure}[!hbtp] 
\center
\resizebox{1\columnwidth}{!}{\rotatebox{0}{\includegraphics{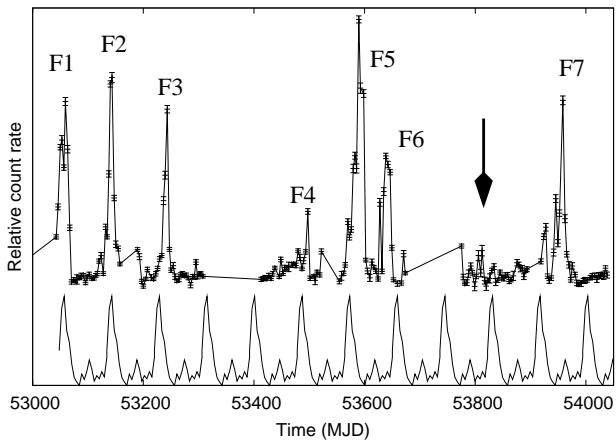}}}
\caption{The PCA monitoring observation lightcurve of Terzan~2 in the
time interval 53013--54050 MJD (\textit{above}) versus a series of
folded light curves with period of 90.55 days (\textit{below}). The
black arrow shows the position of a possible missed flare (see
Section~\ref{sec:lightcurve} }
\label{fig:folded}
\end{figure}

As shown in Figure~\ref{fig:lc}, the flares seem to occur every
$\sim60-100$ days. Both Lomb-Scargle and the PDM techniques confirm
this with a significant signal of period $P\sim90.55$ days.  In
Figure~\ref{fig:folded} we show the PCA lightcurve (top) versus a
20-bin 90.55 days period folded lightcurve (bottom). The folded
lightcurve matches the occurrence of most of the flares. However, it
is clear that the flares are not strictly periodic. For example, F3
seems to occur later and F6 occur earlier than expected. Furthermore,
it is not possible to say if F7 is an early or late flare, or even a
blend of two flares (we observed a small flare which peaked at
$\simeq53926$~MJD, followed by a big one which peaked at
$\simeq53958$~MJD).  Although there are gaps in the data,
Figure~\ref{fig:folded} suggests that some flares do not occur at all
(see arrow in this figure).

\section{Discussion}\label{sec:discussion}

\subsection{Contamination by a second source in the same field of view?}\label{sec:samesource}

As shown in Figure~\ref{fig:lc}, the luminosity of the source slowly
decreases with time during the quiet period 51214--52945 MJD. Although
the rms amplitude changes up to 30\% (see Figure~\ref{fig:rms1000}),
the X-ray timing characteristics are very similar (see Interval A to L
in Figure~\ref{fig:pds}). During the 53000--53700 MJD period, the
source shows flares which show different X-ray timing characteristics
consistent with the island and banana states observed in other atoll
sources \citep[see
e.g.][]{Straaten03,Straaten05,Belloni02,Altamirano05,Altamirano08}.
A possible mechanism of the observed flux variations in Terzan~2 could
be the emergence of a second X-ray source in this globular
cluster unresolved by the $1^{\circ}$ (FWHM) field of view of the PCA. 
If two sources are observed simultaneously with RXTE, then
we would expect to see power spectra which are a combination of the
intrinsic time variability of both sources. 

To further investigate this, we compared the absolute rms amplitude
that we observe both in the quiet and flaring states.
Observation 80105-10-01-00 is the last observation performed during
the quiet period from which we measured an average source countrate of
$\sim200$ counts / second. The integrated power between $7.8 \cdot
10^{-3}$ and $1$~Hz is $(3.4\pm2) 10^{-2}$, which corresponds to a
fractional rms amplitude of $\sim18\pm0.6$\%, i.e. an absolute rms
amplitude of $36\pm1$ counts/second.

Observation 80138-06-01-00 is the first observation performed during
the flaring state. Its average countrate is $\sim740$ counts / second
and the absolute rms amplitude in the $0.0078-1$~Hz range is
$21.4\pm4.4$ counts/second ($2.9\pm0.6$\% fractional rms amplitude).
Clearly, the absolute rms amplitudes are different when comparing
quiet and flaring periods.  If we repeat the analysis using the second
RXTE observation during the flaring state (80138-06-02-00), the
discrepancy is higher. This observation has an average source
countrate of $\sim405$ counts/second and the upper limit for the
absolute rms amplitude in the $7.8 10^{-3}-1$~Hz frequency range is 5
counts/second.

Given the characteristics of the power spectra, flares cannot be
 explained by assuming that another source has emerged, unless
 Terzan~2 turned off at the same time that the other X-ray source
 turned on, which is unlikely. Therefore, we conclude that the flux
 transitions are intrinsic to the only low mass X-ray binary detected
 in the globular cluster Terzan~2: 1E~1724--30 \citep{Revnivtsev02}.

\subsection{The kilohertz QPOs, different states and their transitions}\label{sec:kHz2}

The results presented in this paper show that the low luminosity
source 1E~1724--3045 in the globular cluster Terzan~2 can be
identified as an atoll source. This is the first time a source
previously classified as weak burst source \citep[see e.g,][ and
references within]{Belloni02,Straaten03} showed other states than
those of the extreme island state, confirming previous suggestions
that these sources are atoll sources. We have identified the new
states as the island and banana states based on comparisons between
color color diagrams of different sources and the characteristics of
the power spectra. We have detected at least one of the the kHz QPOs,
and as explained in
Sections~\ref{sec:flaringsegment}~and~\ref{sec:kHz}, in 5 cases we may
be detecting the lower kHz QPO (intervals M and N -- see also
Table~\ref{table:kHz}) and in one case the upper one (interval Q).
No simultaneous twin kHz QPOs were detected within any of the 14
observations that sample the flares. Future observations of flares
will allow us to confirm these identifications and might allow us to
detect both kHz QPOs simultaneously.

We found that the frequencies of the various components in the power
spectra of Terzan~2 followed previously reported relations
(Figures~\ref{fig:nuvsnu}~and~~\ref{fig:pbk}).  Terzan~2 is a
particularly important source in the context of these frequency
correlations because it is one of the few neutron star sources
that has been demonstrated to show power spectral features that reach
frequencies as low as $\simeq0.1$~Hz, which is uncommon for
neutron star low mass X-ray binaries, but not for black holes.
%
Our results demonstrate that in each of the flares, Terzan~2 undergoes
flux transitions that, if directly observed, would probably allow us
to resolve current ambiguities in the identification of components,
such as the case of $L_{\ell ow}$ component in atoll sources. This
component is interpreted by some authors as a broad lower kHz QPO at
very low frequencies \citep[see][]{Psaltis99,Nowak00,Belloni02} which
becomes peaked at higher frequencies, while other authors interpret
$L_{\ell}$ and $L_{\ell ow}$ as different components \citep[see
e.g. discussion in ][]{Straaten03}. Another example is the
identification of the upper kHz QPO at low
frequencies. \citet{Straaten03} have suggested that the broad
component observed at $\simeq150$~Hz in the EIS of atoll sources
becomes the peaked upper kHz QPO $L_u$. These authors based their
interpretation on the frequency correlations shown in
Figure~\ref{fig:nuvsnu}. Nevertheless, as \citet{Straaten03} argue,
these identifications should be taken as tentative.  One way to
confirm the link between them would be to observe the gradual
transformation from one to another one. 

During the time between flares, the source shows intensities similar
to those measured before the quasi-periodic flares started.
Unfortunately there are no observations during those intervals, but we
expect that then Terzan~2 shows X-ray variability similar to that
reported in intervals A--L.  If this is the case, the state transition
between the extreme island state and the island state should be
observable in observations at the beginning or at the end of each
flare. Given the relatively gradual and predictable transitions,
Terzan~2 becomes the best source known up to now to study these
important transitions.

\subsection{On the $\sim90$ days flare recurrence }\label{sec:lightcurve}

The quasi-periodic variations over days and months observed in some
LMXBs X-ray light-curves are generally associated with the possible
precession period of a tilted accretion disk or alternatively long
term periodic variations in the accretion rate or periodic outbursts
of X-ray transients.
Some examples are the $\simeq35$ cycle in Her~X-1 which is thought to
be caused by a varying obscuration of the neutron star by a
tilted-twisted precessing accretion disc; the $\simeq170$ days
accretion cycle of the atoll source 4U~1820--30,
\citep{Priedhorsky84,Simon03}; the 122-125 day cycle in the outbursts
of the recurrent transient Aql X-1 \citep{Priedhorsky84b,Kitamoto93}.
Understanding the mechanisms that trigger the long-term variability
associated with variations in the accretion rate of LMXBs can allow us
to better predict, within each source, when the state transitions
occur. This is useful because these transitions are usually
fast and therefore difficult to observe.

The power spectra of our observations of Terzan~2 during the flaring
confirm that the source undergoes EIS-IS-LLB-LB-UB state transitions,
as observed in other neutron star atoll systems \citep[and not as seen
for Z-sources, see reviews by][ and references
within]{Vanderklis04,Vanderklis06}. As the source increased in X-ray
luminosity, we found that the components in the power spectra
increased in frequency which is consistent with the interpretation
that the accretion disk is moving inwards toward the compact
object. Therefore, the flaring with average 90~days period is most
probably an accretion cycle.
We note that the modulation of the light-curve could be related to the
orbital period of the system or set by the precession of a tilted
disk.  However, the mechanisms involved in those interpretations are
very unlikely to affect the frequency of the kHz QPOs.

\begin{figure}[!hbtp] 
\center
\resizebox{1\columnwidth}{!}{\rotatebox{0}{\includegraphics{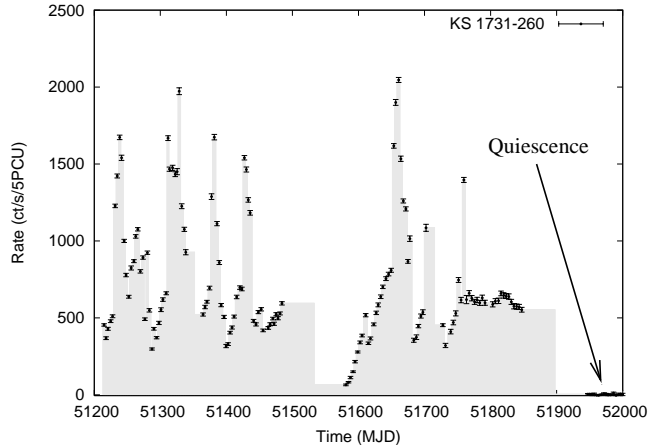}}}
\caption{The PCA monitoring observation lightcurve of the atoll source
  KS~1731--260 during part of its low $L_x$ period. Unfortunately,
  there are no PCA monitoring observations of the source before to
  MJD~51200. Clearly, the source flares up similarly to Terzan~2
  before it turns into quiescence. Interestingly, the data at MJD
  $\sim51550$ shows that the source had a period of very low
  intensity, followed by a flaring up that lasted for $\sim250$~days
  before the source finally turned into quiescence.  }
\label{fig:ks1731}
\end{figure}

If the flares are explained as an accretion cycle, then it is puzzling
why the source underwent a smooth decrease of $L_x$ for
$\simeq8$~years before it started to show the flares. Terzan~2 may not
be the only source that shows this kind of behavior. For example,
KS~1731--260 is a low-luminosity burster that has shown a high $L_x$
phase, during which \citet{Revnivtsev03} reported a possible
$\simeq38$ days period, and a low $L_x$ phase, during which much
stronger variability was observed (which was described as red
noise). After its low $L_x$ phase, KS~1731--260 has turned into
quiescence \citep{Wijnands02a,Wijnands02b}. In Figure~\ref{fig:ks1731}
we show the bulge scan light curve of the source during the low $L_x$
phase. At MJD~$\sim51550$ the source reached very low intensities,
then flared up again for $\gtrsim 250$~days to finally turn into
quiescence. The low luminosities are confirmed by the ASM light curve
(not plotted).
Recently, \citet{Shih05} reported that
the persistent atoll source 4U~1636--53 has also shown a period of
high $L_x$ followed by a period of low $L_x$. During high $L_x$, no
long-term periodicity was found, but a highly significant
$\simeq46$~days period was observed after its $L_x$ decline.

These  similar patterns of behavior 
might point towards a common mechanism, which then must be unaffected
by the intrinsic differences between these sources.

For example, while Terzan~2 remained with approximately constant
luminosity in its extreme island state for $\simeq8$ years before
showing long term periodicities, 4U~1636--53 and KS~1731--260 were
observed with variable luminosity and in different states, including
the banana state in which the kHz QPOs were found (see
e.g. \citealt{Wijnands97b} and \citealt{Shih05} for 4U~1636--53 and
\citealt{Wijnands97a} and \citealt{Revnivtsev03} for
KS~1731--260). While Terzan~2 reached a maximum luminosity of
$L_x/L_{Edd}\simeq0.02$ during one of the flares,
4U~1636--53 shows similar luminosities only at its lowest $L_x$ levels
\citep[while it has reached $L_x/L_{Edd} \gtrsim0.15$ --
see][]{Altamirano08}.  Further differences may be related to whether
these systems are normal or ultra-compact binaries. While 4U~1636--53
is not ultra-compact (see below), \cite{Intzand07} has recently
proposed that Terzan~2 may be classified as ultra-compact based on
measurements of its persistent flux, long burst recurrence times and
the hard X-ray spectra.
If the luminosity behavior of these sources is related, the
differences outlined above suggest that the mechanism that triggers
the modulation of the light curve at low $L_x$ may not depend on the
accretion history, the luminosity of the source or even whether the
system is ultra-compact or not.  The modulation period may depend on
these factors.

Unfortunately, we cannot compare the orbital periods and the
companions of the three systems, as these are only known for
4U~1636--53 \citep[$\sim3.8$~hours and
$\simeq0.4$~M$_{\odot}$][]{Casares06}. Nevertheless, with the present
data it is already possible to exclude some mechanisms.
For example, mass transfer feedback induced by X-ray irradiation
\citep{Osaki85} is unlikely.  In this model, X-ray radiation from the
compact object heats the companion star surface, causing enhancement
of the mass accretion rate in a runaway instability. However, in
\Citeauthor{Osaki85}'s scenario, it is not clear how the system could
remember the phase of the cycle if one of the flares is missed or if
the size of the flares differs much.  Flares F4 and F5 in Terzan~2,
independently of the other two sources, may already raise an objection
to this model.  Although we miss part of F4 due to a gap in the data,
Figure~\ref{fig:lc} shows that F4 was quite short (less than 9 days),
while F5 was the longest ($\lesssim36$~days) and strongest flare.

\citet{Shih05} have suggested that the atoll source 4U~1636--53 may
turn into quiescence after its low $L_x$ period, as was observed for
KS~1731--260. Such an observation for 4U~1636--53 as well as for
Terzan~2 would give credibility to the link between these sources.  To
our knowledge, there is no model which predicts such behavior.

\subsection{Energy dependence as a tool for kHz QPO identification}

\citet{Homan00} discovered a single 695~Hz QPO in the low mass X-ray
binary EXO~0748--676 and identified this QPO as the lower kHz
QPO. These authors based their identification on the fact that at that
time: (i) from the 11 kilohertz QPO pairs found in atoll sources, eight
had ranges of lower peak frequencies that include 695 Hz, which was the
case for only three of the upper peaks and (ii) the upper peaks in atoll
sources generally had Q lower than $\sim14$, while their QPO had
Q$\gtrsim38$, value more common for lower peaks.
While from Figure~\ref{fig:nuvsnu} it can be seen that (i) is not
valid anymore, since the upper kHz QPOs have been detected down to
300--400~Hz (and possibly down to $\lesssim100$~Hz -- see
Section~\ref{pdsqp}),
at these low frequencies $L_u$ is usually much broader than they
observed, which confirms their identification (see
Section~\ref{sec:tfp} and \citealt{Barret05a,Barret05b,Barret05c}).

\citet{Homan00} also based their identification on the comparison of
the energy dependence of the QPO with that of the two kilohertz peaks
in 4U~1608--52, which have rather different energy dependencies
\citep[the power-law rms-energy relation for $L_{\ell}$ is steeper
  than that for $L_u$, see 
][]{Berger96,Mendez98a,Mendez01}. Similarly, \citet{Mendez01} use the
same method to strengthen the identification of the single kHz QPO
observed in the atoll source Aql~X-1. To further investigate if this
method could be used to identify the sharp kHz QPOs we report in
Section~\ref{sec:kHz}, in Figure~\ref{fig:rmsvse} we compare the
energy dependence of the kHz QPOs in 4U~1608--52
\citep{Berger96,Mendez98a,Mendez01} and 4U~0614+09 \citep{Mendez97}
with that of Terzan~2.  The data for Terzan~2 seem to fall in between
those for $L_{\ell}$ and $L_u$ of 4U~1608--52 but shows a completely
different behavior than the data of 4U~0614+09.
The fact that the rms amplitude of the upper kHz QPO in 4U~0614+09 and
4U~1608--52 are significantly different (by up to a factor of 3) and
that the data for Terzan~2 fall in between those of $L_{\ell}$ and
$L_u$ in 4U~1608--52 show that the method does not lead to unambiguous
results. Mean source luminosity, instantaneous luminosity and
instantaneous QPO frequency may all affect QPO energy dependence in
addition to QPO type.

\section{Summary}

\begin{enumerate}

\item We presented a detailed study of the time variability of the
atoll source 1E~1724--3045 (Terzan~2) which includes, for the first
time, observations of this source in its island and banana states
confirming the atoll nature of this source.  We find that the
different states of Terzan~2 show timing behavior similar to that seen
in other NS-LMXBs. Our results for the extreme island state are
consistent with those previously reported in \citet{Belloni02} and
\citet{Straaten03}.

\item We report the discovery of kilohertz quasi-periodic oscillations
(kHz QPOs). Although we do not detect two kHz QPOs simultaneously nor
significant variability above 800~Hz, the detection of the lower and
the upper kHz QPOs at different epochs and the power excess found at
high frequencies (such as the case in intervals~O or observation
90058-06-04-00) suggest that simultaneous twin kHz phenomena as well
as significant variability up to $\sim1100$~Hz (or more) is 
probable.

\item By comparing the dependence of the rms amplitude with energy of
kHz QPOs in the atoll sources 4U~1608--52, 4U~0614+09 and Terzan~2, we
show that this dependence appears to differ between sources and
therefore cannot be used to unambiguously identify the kilohertz QPOs
in either $L_u$ or $L_{\ell}$, as previously thought.

\item We studied the flux transitions or flares observed since
February 2004 and from the source state changes observed we conclude
that they are due to aperiodic changes in the accretion rate.

\item State transitions between the extreme island state and the
island state should be observable in observations at the beginning or
at the end of each flare. Given the relatively gradual and predictable
transitions, Terzan~2 becomes the best source known upto now to study
such transitions.

\end{enumerate}

\center
\begin{longtable}{|c|c|c|c|}
\hline
Characteristic & Q & rms & Component \\
Frequency (Hz) &   & (\%) & ID \\
\hline
\endfirsthead
\hline
Characteristic & Q & rms & Component \\
Frequency (Hz) &   & (\%) & ID \\
\hline
\endhead
\hline \multicolumn{4}{|r|}{{Continued on next page}} \\ \hline
\endfoot
\hline \hline
\endlastfoot
\hline \multicolumn{4}{|c|}{Interval A} \\ \hline
$   156.9\pm      6.6$ & $     0.25\pm      0.06$ & $    13.9\pm      0.4$ &$L_{u}$\\ 
$    11.5\pm      0.2$ & $     0.04\pm      0.03$ & $    17.6\pm      0.3$ &$L_{low}$\\ 
$     1.05\pm      0.01$ & $     0.47\pm      0.02$ & $    16.3\pm      0.3$ &$L_{h}$\\ 
$     0.820\pm      0.006$ & $     5.76\pm      0.87$ & $     3.3\pm      0.3$ &$L_{LF}$\\ 
$     0.171\pm      0.005$ & $     0.32\pm      0.06$ & $    12.4\pm      0.6$ &$L_{b}$\\ 
\hline \multicolumn{4}{|c|}{Interval B} \\ \hline
$   200.1\pm     11.6$ & $     0.55\pm      0.09$ & $    12.6\pm      0.5$ &$L_{u}$\\ 
$    19.1\pm      1.5$ & $     --$                & $    13.2\pm      0.2$ &$L_{low}$\\ 
$     2.59\pm      0.05$ & $     0.40\pm      0.04$ & $    13.6\pm      0.4$ &$L_{h}$\\ 
$     0.40\pm      0.02$ & $     --$              & $    12.2\pm      0.2$ &$L_{b}$\\ 
\hline \multicolumn{4}{|c|}{Interval C} \\ \hline
$   148.4\pm      6.8$ & $     0.27\pm      0.05$ & $      13.8\pm      0.3$ &$L_{u}$\\ 
$     9.7\pm      0.1$ & $     --$                & $      18.6\pm      0.1$ &$L_{low}$\\ 
$     0.86\pm      0.01$ & $     0.55\pm      0.02$ & $    15.1\pm      0.3$ &$L_{h}$\\ 
$     0.633\pm      0.007$ & $     6.2\pm      1.2$ & $     3.1\pm      0.3$ &$L_{LF}$\\ 
$     0.158\pm      0.006$ & $     --$              & $    14.5\pm      0.3$ &$L_{b}$\\ 
$     0.106\pm      0.006$ & $     1.94\pm      1.02$ & $   3.4^{+0.8}_{-0.5}$ &$L_{vl}$\\ 
\hline \multicolumn{4}{|c|}{Interval D} \\ \hline
$   130.8\pm      7.4$ & $     0.21\pm      0.07$ & $    13.1\pm      0.3$ &$L_{u}$\\ 
$     8.5\pm      0.20$ & $     --$               & $    18.5\pm      0.1$ &$L_{low}$\\ 
$     0.74\pm      0.01$ & $     0.53\pm      0.03$ & $  14.5\pm      0.3$ &$L_{h}$\\ 
$     0.512\pm      0.004$ & $     6.5\pm      0.9$ & $   3.9\pm      0.2$ &$L_{LF}$\\ 
$     0.131\pm      0.005$ & $     --$              & $  15.4\pm      0.2$ &$L_{b}$\\ 
\hline \multicolumn{4}{|c|}{Interval E} \\ \hline
$   169.2\pm      8.6$ & $     0.31\pm      0.07$ & $    13.6\pm      0.4$ &$L_{u}$\\ 
$    10.8\pm      0.2$ & $     0.01\pm      0.04$ & $    18.2\pm      0.3$ &$L_{low}$\\ 
$     0.92\pm      0.01$ & $     0.45\pm      0.03$ & $    16.6\pm      0.3$ &$L_{h}$\\ 
$     0.750\pm      0.006$ & $     7.8\pm      1.8$ & $     2.8\pm      0.2$ &$L_{LF}$\\ 
$     0.144\pm      0.004$ & $     0.34\pm      0.07$ & $    12.7\pm      0.6$ &$L_{b}$\\ 
\hline \multicolumn{4}{|c|}{Interval F} \\ \hline
$   216.1\pm      2.2$ & $    17.9\pm      8.4$ & $     3.1\pm      0.5$ &$***$\\ 
$   161.4\pm     17.0$ & $     0.14\pm      0.13$ & $    14.2\pm      0.8$ &$L_{u}$\\ 
$    12.1\pm      0.5$ & $     0.10\pm      0.07$ & $    16.6\pm      0.6$ &$L_{low}$\\ 
$     1.21\pm      0.04$ & $     0.44\pm      0.04$ & $    15.6\pm      0.5$ &$L_{h}$\\ 
$     0.93\pm      0.01$ & $     3.5\pm      0.9$ & $     4.4\pm      0.6$ &$L_{LF}$\\ 
$     0.179\pm      0.009$ & $     0.09\pm      0.04$ & $    13.8\pm      0.3$ &$L_{b}$\\ 
\hline \multicolumn{4}{|c|}{Interval G} \\ \hline
$    17.5\pm      2.4$ & $     --$                  & $    19.6\pm      0.5$ &$L_{low}$\\ 
$     1.17\pm      0.07$ & $     0.3\pm      0.1$ & $    16.4\pm      1.1$ &$L_{h}$\\ 
$     0.716\pm      0.008$ & $    11.2^{24.5}_{3.6} $ & $     4.1\pm      0.6$ &$L_{LF}$\\ 
$     0.152\pm      0.016$ & $     --$                  &  $    16.4\pm      0.7$ &$L_{b}$\\ 
\hline \multicolumn{4}{|c|}{Interval H} \\ \hline
$   131.7\pm     15.8$ & $     0.3\pm      0.1$ & $    13.2\pm      0.8$ &$L_{u}$\\ 
$     9.6\pm      0.4$ & $     --$                  & $    18.5\pm      0.2$ &$L_{low}$\\ 
$     0.892\pm      0.026$ & $     0.47\pm      0.04$ & $    14.5\pm      0.4$ &$L_{h}$\\ 
$     0.634\pm      0.006$ & $     4.6\pm      0.8$ & $     4.5\pm      0.4$ &$L_{LF}$\\ 
$     0.133\pm      0.005$ & $     --$                  & $    16.3\pm      0.2$ &$L_{b}$\\ 
\hline \multicolumn{4}{|c|}{Interval I} \\ \hline
$   211.7\pm     40.3$ & $     --$                  & $    19.4\pm      0.9$ &$L_{u}$\\ 
$     9.05\pm      0.39$ & $     0.17\pm      0.07$ & $    18.8\pm      0.6$ &$L_{low}$\\ 
$     0.87\pm      0.02$ & $     0.47\pm      0.06$ & $    17.6\pm      0.7$ &$L_{h}$\\ 
$     0.67\pm      0.01$ & $     8.2\pm      3.3$ & $     3.2\pm      0.4$ &$L_{LF}$\\ 
$     0.136\pm      0.009$ & $     0.13\pm      0.05$ & $    16.2\pm      0.5$ &$L_{b}$\\ 
\hline \multicolumn{4}{|c|}{Interval J} \\ \hline
$    86.4\pm     32.8$ & $     --$                  & $    14.7\pm      1.6$ &$L_{u}$\\ 
$     9.71\pm      1.07$ & $     0.19\pm      0.13$ & $    17.3\pm      1.5$ &$L_{low}$\\ 
$     0.8\pm      0.1$ & $     --$                  & $    17.6\pm      1.6$ &$L_{h}$\\ 
$     0.60\pm      0.02$ & $     1.5\pm      0.4$ & $     8.8^{+2.2}_{-1.2}$ &$L_{LF}$\\ 
$     0.097\pm      0.008$ & $     0.27\pm      0.08$ & $    14.4\pm      0.9$ &$L_{b}$\\ 
\hline \multicolumn{4}{|c|}{Interval K} \\ \hline
$   122.5\pm     18.6$ & $     0.04\pm      0.3$ & $    16.7\pm      1.2$ &$L_{u}$\\ 
$     7.3\pm      0.5$ & $     --$                  & $    18.8\pm      0.4$ &$L_{low}$\\ 
$     0.79\pm      0.03$ & $     0.54\pm      0.06$ & $    14.0\pm      0.6$ &$L_{h}$\\ 
$     0.556\pm      0.006$ & $     5.1\pm      1.5$ & $     4.3\pm      0.5$ &$L_{LF}$\\ 
$     0.134\pm      0.006$ & $     --$                  & $    17.2\pm      0.3$ &$L_{b}$\\ 
\hline \multicolumn{4}{|c|}{Interval L} \\ \hline
$   202.7\pm     94.6$ & $     --$ & $    13.04\pm      2.08$ &$L_{u}$\\ 
$    16.3\pm      1.8$ & $     --$ & $    18.3\pm      0.6$ &$L_{low}$\\ 
$     1.34\pm      0.05$ & $     0.48\pm      0.07$ & $    15.5\pm      0.6$ &$L_{h}$\\ 
$     0.98\pm      0.01$ & $    10.06\pm      8.60$ & $     3.3^{+0.6}_{-0.5}$ &$L_{LF}$\\ 
$     0.19\pm      0.01$ & $     --$ & $    16.1\pm      0.4$ &$L_{b}$\\ 
\hline \multicolumn{4}{|c|}{Interval M} \\ \hline
$   758.9\pm      4.8$ & $    12.2^{+12}_{-4}$ & $     9.1\pm      0.7$ &$L_{\ell}$\\ 
$    39.4\pm      2.5$ & $     1.3\pm      0.6$ & $     6.4\pm      0.6$ &$L_{b}$\\ 
\hline \multicolumn{4}{|c|}{Interval N} \\ \hline
$   586.1\pm      6.6$ & $     5.5\pm      1.24$ & $     10.6\pm      0.8$ &$L_{\ell}$\\ 
$    29.1\pm      1.1$ & $     1.1\pm      0.2$ & $     8.4\pm      0.5$ &$L_{b}$\\ 
$     4.9\pm      1.6$ & $     0.47\pm      0.3$ & $     3.9^{+0.8}_{-0.5}$ &$L_{b2}$\\ 
\hline \multicolumn{4}{|c|}{Interval O} \\ \hline
$    30.5\pm      6.3$ & $     0.28\pm      0.17$ & $     5.8\pm      0.4$ &$***$\\ 
$     0.025\pm      0.002$ & $     0.3\pm      0.1$ & $     3.3\pm      0.1$ &$***$\\ 
\hline \multicolumn{4}{|c|}{Interval P} \\ \hline
$    30.9\pm      0.4$ & $     2.02\pm      0.21$ & $    12.6\pm      0.6$ &$***$\\ 
$    15.5\pm      1.6$ & $     1.005\pm      0.156$ & $     7.9\pm      0.9$ &$***$\\ 
\hline \multicolumn{4}{|c|}{Interval Q} \\ \hline
$   565.0\pm     24.5$ & $     2.4\pm      0.6$ & $    15.10\pm      1.44$ &$L_{u}$\\ 
$   148.7\pm     23.3$ & $     0.7\pm      0.4$ & $    12.4\pm      1.8$ &$L_{hHz}$\\ 
$    28.02\pm      2.37$ & $     1.06\pm      0.56$ & $     9.4^{+2.7}_{-1.8}$ &$L_{h} ^{ **} $\\ 
$     8.1\pm      1.4$ & $     0.10\pm      0.08$ & $    15.1\pm      1.2$ &$L_{b}$\\ 
\hline \multicolumn{4}{|c|}{Interval R} \\ \hline
$    68.5\pm      7.2$ & $     1.8\pm      1.1$ & $     5.5\pm      0.9$ &$***$\\ 
\hline
\caption{Characteristic frequencies $\nu_{max}$, $Q$ values
($\nu_{max} \equiv \nu_{central}/FWHM$), fractional rms (in the full
PCA energy band) and component identification (ID) of the Lorentzians
fitted for Terzan~2.  The quoted errors use $\Delta\chi^2 =
1.0$. Where only one error is quoted, it is the straight average
between the positive and the negative error. ``$--$'': Broad
Lorentzians whose quality factor was fixed to 0. \\ $^{**}$: This
component is not $3\sigma$ significant. However, without this
component the fit becomes unstable (see Section~\ref{sec:gralpds}). \\ $***$:
For this component, no clear identification was possible. }
\label{table:data}
\end{longtable}


\clearpage


\end{document}